\documentclass[letter,preprintnumbers,superscriptaddress,nofootinbib]{revtex4}
\usepackage{hyperref}
\hypersetup{colorlinks=true -,linkcolor=purple,anchorcolor=blue,citecolor=blue, filecolor=blue,urlcolor=red,bookmarksnumbered=true,
	pdfview=FitB
}
\usepackage{subcaption}
\usepackage{color}
\usepackage{xcolor}
\usepackage{ulem}
\usepackage{booktabs}

\usepackage{csquotes}
\colorlet{purple1}{blue!70!red}
\colorlet{darkred}{red!50!black}
\usepackage{graphicx}
\usepackage{psfrag}
\usepackage{color}
\usepackage{comment}
\usepackage{amssymb}
\usepackage{amsmath}
\usepackage{epstopdf}
\usepackage{epsf}
\def\be{\begin{equation}}
\def\ee{\end{equation}}
\def\ba{\begin{array}}
\def\ea{\end{array}}
\def\bc{\begin{center}}
\def\ec{\end{center}}

\newcommand{\eq}{\begin{eqnarray}}
\newcommand{\en}{\end{eqnarray}}
    
\newcommand{\PC}{\color{orange}}
\def \p {\partial}

\def \to {\rightarrow}

\def \pn3 {\psi_{u\bar d g}}

\def \pu4 {\psi_{u\bar d q\bar q}}

\def \p4n {\psi_{u\bar d gg}}

\def \pp4 {\psi_{uudg}}

\date{\today}

\usepackage{xcolor}
\usepackage{tikz-feynman}
\usepackage{soul}

\usepackage{natbib}

\begin{document}
\title{Spin–Orbit Correlations in the Pion and the Role of Quark-gluon Interaction}
   \author{Poonam~Choudhary}
    \email{poonam.hep.ph@iitb.ac.in} 
\author{Asmita~Mukherjee}
	\email{asmita@phy.iitb.ac.in} 
 \author{Ravi~Singh}
	\email{ravi22998singh@gmail.com} 
	\affiliation{Department of Physics, Indian Institute of Technology Bombay, Powai, Mumbai 400076, India}	
\date{\today}

\begin{abstract}
We study  the {{spin-orbit correlations (SOCs)  of the pion using overlap of light-front wave functions (LFWFs). Going beyond the leading Fock sector, we  incorporate one gluon in the wave function. The analytic form of the higher Fock component of the LFWF is}} constructed by incorporating a perturbative gluon to the pion state. 
{{This allows us to explore the role of quark-gluon interactions in the spin-orbit correlation within a model calculation.}} We investigate the kinetic and canonical spin–orbit correlations of quarks in the pion, {{which  arise from different decompositions of the energy-momentum tensor.}}
 We further explore {the difference between kinetic and canonical SOC arising from the inclusion of higher Fock sector containing gluon.}
 

\end{abstract}
\date{\today}
\maketitle

\section{Introduction}
\label{s:intro}
The internal spin structure of hadrons \cite{Aidala:2012mv,COMPASS:2007rjf,Bass:2004xa,Ermolaev:2025axz,Leader:2013jra,Ji:1996ek} remains one of the central puzzles in QCD.  In particular, the pion, as the lightest meson and a pseudo-Goldstone boson \cite{Nambu:1961tp,Roberts:2020hiw}, offers a unique laboratory.  Its spin is zero, yet the dynamics of its constituents  quarks, antiquarks, gluons and sea quarks  encode rich information about orbital motion, spin correlations and the role of quark gluon interactions\cite{deFlorian:2014yva,Chang:2014lva}.
The study of correlations among partonic degrees of freedom such as spin, momentum and position has opened a new window into the multidimensional structure of hadrons \cite{Diehl:2003ny,Belitsky:2005qn,Ji:2004gf,Rajan:2016tlg}. These correlations go beyond one-dimensional parton distributions and encode the interplay between intrinsic spin, orbital motion, and color interactions inside hadrons. They form the basis of our present understanding of how confined partons generate the total mass, momentum, spin and angular momentum of a hadron \cite{Ji:1996ek,Leader:2013jra,Burkardt:2012sd}. A well-known example is the spin-momentum correlation that gives rise to the Sivers function~\cite{Sivers:1990fh,Brodsky:2002cx,Collins:2002kn}, which describes the asymmetric distribution of unpolarized quarks in a transversely polarized nucleon. Similar correlations between the quark’s transverse spin and its transverse momentum lead to the Boer Mulders function~\cite{Boer:1997nt,Bacchetta:2004jz}. These effects observed in semi-inclusive deep inelastic scattering (SIDIS)~\cite{Bacchetta:2006tn,Ji:2004wu} and Drell–Yan processes, reflect how parton motion and color interactions create measurable spin asymmetries.  
On the other hand, spin-spin correlations are studied through helicity and transversity distributions ~\cite{Barone:2001sp,Anselmino:2007fs}, tensor charges, and polarization observables, revealing how quark spins align or anti-align with respect to the hadron spin. 
Another class of correlations links the quark spin to its orbital angular momentum (OAM), known as spin–orbit correlations (SOCs) \cite{Bhattacharya:2024sno, Lorce:2014mxa, Lorce:2025ayr,Tan:2021osk,Engelhardt:2021kdo}. {Although spin–orbit correlations do not enter the total spin or angular-momentum decomposition as independent contributions but they characterize how the quark’s spin and orbital angular momentum are intertwined and provides essential information about the internal dynamics of the hadron. It is analogous to spin-orbit coupling of an electron in the hydrogen atom which contributes to its fine structure. In fact in ~\cite{Hatta:2024otc}, Hatta et al. derived a new momentum sum rule which includes both quark and gluon SOC.}
The concept of decomposing the total angular momentum of hadrons into its quark and gluon components has evolved through several theoretical frameworks \cite{Leader:2013jra,Chen:2008ag,Wakamatsu:2014zza}. In the Ji decomposition~\cite{Ji:1996ek}, the quark angular momentum and SOC are expressed in terms of gauge–invariant local operators involving the covariant derivative.  
This definition naturally leads to what is known as the {kinetic} decomposition, which incorporates the effects of the gluon field through the gauge connection.  
The Ji sum rule links the total kinetic OAM and SOC to measurable quantities, specifically to the moments of generalized parton distributions (GPDs). 
This formalism provided one of the first gauge–invariant ways to relate partonic angular momentum to experimental observables. In contrast, Jaffe and Manohar~\cite{Jaffe:1989jz} introduced an alternative decomposition where the quark OAM and SOC are defined using the ordinary derivative rather than the covariant one. This version is referred to as the {canonical} decomposition and it omits explicit gluon-field terms and thus depends on the choice of gauge. While it captures the intuitive partonic picture of orbital motion, it is not gauge invariant in general. Lorcé and collaborators~\cite{Lorce:2011kd,Lorce:2017wkb} further connected these operator definitions to generalized parton distributions (GPDs) and generalized transverse–momentum dependent distributions (GTMDs), which describe the complete phase-space correlations between quark momentum, position, and spin. 

Hatta~\cite{Hatta:2011ku,Hatta:2012cs} later provided a gauge-invariant extension of the Jaffe–Manohar framework by introducing Wilson lines that encode the color flow generated by initial- and final-state interactions. This reformulation clarified that the difference between the canonical and kinetic OAM corresponds to the potential angular momentum \cite{Hatta:2011ku,Wakamatsu:2010qj, Burkardt:2012sd}, which reflects the torque exerted by the color–Lorentz force of the gluon field on the active parton. 
Analogously, the difference between the canonical and kinetic SOCs defines the potential SOC, which has an identical physical origin in the color-induced torque \cite{Hatta:2024otc}. Although this quantity is less explored in the literature, it plays an important role in understanding how quark spin couples to quark OAM in the presence of gauge-field interactions. While these ideas were primarily developed for the nucleon, their extension to the pion provides new insight into how spin–orbit structure can appear in a spinless hadron. In the pion, both the canonical and kinetic OAM vanish due to its spin zero nature. 
However, the spin–orbit correlation remains nonzero in both the kinetic and canonical formulations.
In ~\cite{Hatta:2024otc}, the authors argued that the difference between kinetic and canonical SOC is directly proportional to the twist-3 quark-gluon correlation function. Once gluonic degrees of freedom are introduced, either in the form of higher twist quark-gluon correlation or through higher Fock states with gluons, the canonical and kinetic SOCs show difference which reflects the {affect} of color Lorentz force acting on quarks.

Traditionally, the model studies have focused on the leading Fock sector of the pion, often neglecting the contributions from gluons and higher Fock sectors. Such approximations are justified at low energy scales where valence quarks dominate, but they overlook important gluonic effects that emerge at higher momentum transfers or under QCD evolution. Recent global QCD analyses of pion parton distribution functions (PDFs) ~\cite{Pasquini:2023aaf,Barry:2018ort,Barry:2021osv} incorporate Drell–Yan, prompt-photon and leading-neutron electroproduction data. These studies have revealed that gluons carry a sizable fraction of the pion’s total momentum at typical hadronic scales~\cite{deFlorian:2014yva,Barry:2018ort,Novikov:2020snp,Barry:2021osv,Chang:2020rdy,Lan:2025qio,Lan:2024ais,Lan:2021wok}. These findings emphasize that gluons play a crucial role in the pion’s partonic structure and cannot be neglected in studies of its spin–orbit correlations (SOCs) and angular momentum decomposition \cite{Bhattacharya:2024sck,Ji:2009fu,Ji:2010zza}.
Tan and Lu~\cite{Tan:2021osk} recently explored the longitudinal spin–orbit correlation in the pion using a light-cone quark model restricted to the valence Fock sector. 
Beyond such models, a variety of theoretical approaches such as Dyson–Schwinger equations~\cite{Chang:2013pq,Mezrag:2014jka}, Bethe–Salpeter frameworks~\cite{Bednar:2018mtf,Shi:2020pqe}, light-front constituent models~\cite{Lan:2019vui,Broniowski:2017wbr,Acharyya:2024enp}, AdS/QCD models ~\cite{Chakrabarti:2023djs,Choudhary:2023unw,Gurjar:2021dyv} and lattice QCD simulations~\cite{PhysRevD.103.014508,Detmold:2021qln} have provided complementary insights into the pion’s quark and gluon distributions, form factors and GPDs. On the experimental side, direct measurements of the spin–orbit correlation in the pion are not yet available. Nevertheless, lattice QCD computations of pion GPD moments, electromagnetic form factors, and charge radii~\cite{QCDSF-UKQCD:2006nvr, PhysRevD.103.014508} provide valuable indirect constraints. These studies collectively highlight the importance of including gluon dynamics to achieve a complete three-dimensional picture of the pion. Recent and upcoming experimental programs at facilities such as Jefferson Lab 12~GeV and the future Electron–Ion Collider (EIC) aim to access the pion’s three-dimensional structure via tagged deep inelastic scattering and deeply virtual Compton scattering (DVCS) ~\cite{Accardi:2023chb,Arrington:2021alx,Anderle:2021wcy,AbdulKhalek:2021gbh,JeffersonLabSoLID:2022iod}. 

In this work, we extend previous model studies by explicitly including a perturbative gluon exchange in the pion’s Fock-state expansion within a light-front framework. This allows us to construct light-front wave functions (LFWFs) beyond the valence sector, incorporating quark–gluon-quark interaction at $\mathcal{O}(\alpha_s)$. Using these extended LFWFs, we explore various structural observables of the pion, including its parton distribution functions (PDFs), generalized parton distributions (GPDs), form factors, and generalized transverse momentum–dependent distributions (GTMDs), along with the associated spin–orbit correlations (SOCs). In particular, we calculate both the canonical and kinetic quark SOCs and examine their difference, which originates from the quark–gluon dynamics. 
This study provides a first step toward quantifying gluon contributions to quark spin–orbit correlations and assessing their implications for hadronic spin structure. 
The results offer useful guidance for future theoretical and lattice studies and may serve as a baseline for upcoming experimental efforts aimed at exploring the pion’s three-dimensional structure.

\section{Quark Orbital Angular Momentum and Spin Orbit Correlation}
\label{SOC}
The matrix element of parton OAM describes the correlation between parton's OAM and hadron spin, whereas the matrix element of SOC describes the correlation between a parton’s intrinsic spin and its orbital motion inside a hadron. 
In QCD, two distinct types of OAM can be defined depending on 
which decomposition of energy momentum tensor (EMT) or generalized angular momentum tensor is chosen: kinetic and canonical OAM 
\cite{Leader:2013jra, Lorce:2017wkb}. Similarly, kinetic and canonical SOC have been defined depending on the type of EMT \cite{Hatta:2024otc}.

\subsection{Kinetic OAM and SOC}
The kinetic OAM naturally appears in the Ji decomposition~\cite{Ji:1996ek} of the total angular momentum operator, which in the light–front formalism can be expressed as
	
	{ \be
	\hat J_{z,\text{kin}}=\hat S^q_{z,\text{kin}}+\hat L^q_{z,\text{kin}}+\hat J^G_{z,\text{kin}}.
	\ee
	The kinetic OAM is related to the angular momentum density by following relation \cite{Lorce:2014mxa,Lorce:2025ayr}: 
	\be
	\hat{L}^q_{z, \text{kin}}= -\frac{1}{2}\varepsilon^{3jk} \int d^3x \hat{M}^{0jk}_{\text{kin}} = \frac{1}{2} \int \mathrm{d}^3 x \,  \, \bar{\psi} \, \gamma^+ \, (\mathbf{x} \times i \overset{\leftrightarrow}{\mathbf{D}})_z \, \psi
= \hat{L}^{q_R}_{z,\text{kin}} + \hat{L}^{q_L}_{z,\text{kin}}, \label{L}
	\ee
where $\hat{L}^{q_{R,L}}_{z,\text{kin}}=\bar{\psi}^{R,L} \, \gamma^+ \, (\mathbf{x} \times i \overset{\leftrightarrow}{\mathbf{D}})_z \, \psi^{R,L}$ are the right and left handed quark contributions to total quark OAM, $\overleftrightarrow{\mathbf{D}}=\overrightarrow{\boldsymbol{\partial}}-\overleftarrow{\boldsymbol{\partial}}-2ig\boldsymbol{A}$ denotes the covariant derivative and the angular momentum density
	$\hat{M}^{\alpha \mu \nu}_{\text{kin}} = x^\mu \hat{T}^{\alpha \nu}_{\text{kin}} - x^\nu \hat{T}^{\alpha \mu}_{\text{kin}} $ is 
	expressed in terms of the energy momentum tensor $\hat{T}^{\mu \nu}_{\text{kin}}$ which is 
    \begin{equation}
		\hat{T}^{\mu \nu}_{\text{kin}}= \tfrac{1}{2}\,\bar{\psi}\,\gamma^\mu \, i\, \overleftrightarrow{\mathbf{D}^\nu }\psi =\hat{T}^{\mu\nu}_{\text{kin}, q_R} + \hat{T}^{\mu\nu}_{\text{kin}, q_L} 
	\end{equation}
	
The matrix element of the quark OAM and spin must be proportional to the spin of the hadron because these matrix elements represent the correlation of quark OAM to hadron spin and quark spin to hadron spin, respectively \cite{Lorce:2015sqe}, both of which are zero for a pion. Thus $L^q_{z,\text{kin}}$ and $S^q_{z,\text{kin}}$ are zero leading to vanishing total angular momentum. }
 This implies that in pion, $ \hat{L}^{q_R}_{z,\text{kin}}=-\hat{L}^{q_L}_{z,\text{kin}}$. However, the kinetic spin orbit correlation operator which is defined as difference of right and left-handed quark contribution ~\cite{Lorce:2014mxa,Lorce:2025ayr}: 
\begin{align}
\hat{C}^q_{\text{kin},z} &= \int \mathrm{d}^3 x \, \frac{1}{2} \, \bar{\psi} \, \gamma^+ \gamma_5 \, (\mathbf{x} \times i \overset{\leftrightarrow}{\mathbf{D}})_z \, \psi 
= \hat{L}^{q_R}_{z,\text{kin}} - \hat{L}^{q_L}_{z,\text{kin}}=-2\hat{L}^{q_L}_{z,\text{kin}}=2 \hat{L}^{q_R}_{z,\text{kin}} \label{C}
\end{align}
is evidently non zero. Interestingly, this difference can also be expressed in terms of the parity–odd energy–momentum tensor and the quark spin–orbit correlation can be rewritten as \cite{Lorce:2025ayr}  
\begin{equation}
		\hat{C}^q_{\text{kin}, z} = \int d^3x \,(x^1 \hat{T}^{+2}_{\text{kin}, q5} - x^2 \hat{T}^{+1}_{\text{kin}, q5}),
		\label{OAM5}
	\end{equation}
where $\hat{T}^{\mu\nu}_{\text{kin}, q5}$ acts as the parity–odd partner of the quark kinetic EMT,  
\begin{align}
		\hat{T}^{\mu\nu}_{\text{kin}, q5} = \tfrac{1}{2}\,\bar{\psi}\,\gamma^\mu \gamma_5\, i\, \overleftrightarrow{\mathbf{D}^\nu }\psi
		= \hat{T}^{\mu\nu}_{\text{kin}, q_R} - \hat{T}^{\mu\nu}_{\text{kin}, q_L} 
			\end{align}
	where $\overleftrightarrow{\mathbf{D}^\nu } = \overrightarrow{\boldsymbol{\partial}^\nu}- \overleftarrow{\boldsymbol{\partial}^\nu}-2ig\boldsymbol{A}^\nu(x)$. The corresponding matrix element of odd EMT tensor is parameterized in terms of Form factors (FFs) \cite{Hagler:2009ni,Lorce:2025ayr} :
	\begin{equation}
		\langle p'|\hat{T}^{\mu\nu}_{\text{kin}, q5}|p \rangle
		= i \epsilon^{\mu \nu \Delta P} \tilde{F}^{q} (t)
		\label{EMTparam}
	\end{equation}
	where $P=\tfrac{1}{2}(p'+p)$, $\Delta=p'-p$, $t=\Delta^2$, and $\epsilon^{\mu \nu \Delta P} \equiv \epsilon^{\mu \nu \alpha \beta}\Delta_\alpha P_\beta$ with $\epsilon_{0123}=1$.  Using the above relation, we can write the expectation value of SOC in terms of the Form factor:
	\begin{equation}
C^q_{\text{kin},z}(t)=-\frac{\epsilon_{mn3}}{2P^{+}}\left[i\,\frac{\partial}{\partial\Delta_m}\langle p^{\prime}|\hat{T}^{+n}_{\text{kin}, q5}(0)|p\rangle\right]=\tilde F^q(t), \label{CqEMT}
	\end{equation}
	where $m=1,2$.
	$\tilde F^q(t)$ is basically coming from vector and tensor current and can be expressed in terms of vector FF and tensor FF of the pion 
	\begin{equation}\label{eom}
		\tilde F^q(t)=\frac{1}{2} \left[-F^q(t)+\frac{m_q}{M}\,H^q(t) \right],
	\end{equation}
where $ F^q(t) $ is analogous to Dirac Form factor of vector current and $ H^q(t)$ is tensor FF. 
These quantities are connected to Generalized Parton Distributions (GPDs) through their moments.
At leading twist for the pion, two independent quark GPDs appear: a chiral-even unpolarized GPD originating from the vector current and a chiral-odd GPD arising from the tensor current \cite{Meissner:2008ay}. 
\begin{align}
	F_1^{\mathcal{\pi}}(x,\xi,t)&= \int \frac{dz^-}{4 \pi} e^{i x P^+ z^-/2}
	\, \langle\mathcal{\pi}(P')\vert\bar{\Psi}_q(0)\gamma^+ \Psi_q(z)\vert\mathcal{\pi}(P)\rangle\vert_{z^+=\textbf{z}^\perp=0},\label{H}\\
	-\frac{i \epsilon_{T}^{i j} \Delta_T^{i}}{M_\pi} H_1^{\mathcal{\pi}}(x,\xi,t)&= \int \frac{dz^-}{4 \pi} e^{i x P^+ z^-/2}
	\,\langle\mathcal{\pi}(P')\vert\bar{\Psi}_q(0)i \sigma^{j+}\gamma_5 \Psi(z)_q\vert\mathcal{\pi}(P)\rangle\vert_{z^+=\textbf{z}^\perp=0},\label{ET}
\end{align} 
We can relate the Form factors  as the  moments of the GPDs \cite{Freese:2019bhb, Tanaka:2018wea,Krutov:2020ewr}, evaluated at either zero or non-zero skewness, $\xi=\dfrac{\Delta^+}{2P^+}$:
\begin{equation}\label{sl-pi1}
	F^q(t)= \int dx\, F^q_1(x, \xi, t), \hspace*{0.5cm}
    A^q(t)= \int dx\, x F^q_1(x, \xi, t),\hspace*{0.5cm}
	H^q(t)= 2\int dx\, H^q_1(x, \xi, t),
\end{equation}
Here $F^q(t)$ denotes the quark electromagnetic (vector) form factor of the pion, 
$A^q(t)$ is the quark contribution to the pion gravitational (energy-momentum) form factor, 
and $H^q(t)$ represents the quark tensor (transversity) form factor of the pion.
Using this connection, the quark kinetic spin–orbit correlation in the pion can be expressed as
\begin{equation}
	C_{\text{kin},z}^{q/\pi}(t) = \frac{1}{2} \int dx\,  \left[- F_1^{\mathcal{\pi}}(x,\xi,t) +\frac{m_q}{M} H_1^{\mathcal{\pi}}(x,\xi,t) \right]  \label{CqGPD}
\end{equation}
{ Thus, }in the forward limit, $C_{\text{kin},z}^{q/\pi}(0)$ {is equal to} minus half of first moment of unpolarized PDF { i.e. minus half of the number of valence quarks in the pion}. This indicates that the correlation between quark spin and orbital motion is also related to quark number density. 
The chiral odd GPD contribution along with $ \frac{m_q}{M}$ factor is suppressed and  can be neglected safely in the forward limit. 
\subsection{Canonical OAM and SOC}
The Jaffe--Manohar (JM) decomposition~\cite{Jaffe:1989jz} is based on the {canonical} energy-momentum tensor.  
From the spatial components one obtains the canonical quark orbital angular momentum (OAM) operator,
\begin{equation}
\hat{L}^q_{z} = \int \mathrm{d}^3 x \, \frac{1}{2} \, \bar{\psi} \, \gamma^+ \, (\mathbf{x} \times i \overset{\leftrightarrow}{\mathbf{\partial}})_z \, \psi
\end{equation}
which involves {ordinary} derivatives and therefore reflects a fully canonical structure. $\hat{L}^q_z$ is also zero for the pion owing to the same reasons as discussed for kinetic quark OAM. This again implies $ \hat{L}^{q_R}_{z}=-\hat{L}^{q_L}_{z}$. We can also write corresponding SOC in terms of canonical EMT: 
\begin{align}
\hat{C}^q_{z} & = \int \mathrm{d}^3 x \, \frac{1}{2} \, \bar{\psi} \, \gamma^+ \gamma_5 \, (\mathbf{x} \times i \overset{\leftrightarrow}{\mathbf{\partial}})_z  \, \psi =\int d^3x \,(x^1 \hat{T}^{+2}_{q5} - x^2 \hat{T}^{+1}_{q5}) \\
&{ = \hat{L}^{q_R}_{z} - \hat{L}^{q_L}_{z}=-2\hat{L}^{q_L}_{z}=2 \hat{L}^{q_R}_{z}}
\end{align}

where 
\begin{align}
    \hat{T}^{\mu\nu}_{q5} &= \tfrac{1}{2}\,\bar{\psi}\,\gamma^\mu \gamma_5\, i\, \overleftrightarrow{\partial^\nu }\psi
		= \hat{T}^{\mu\nu}_{q_R} - \hat{T}^{\mu\nu}_{q_L} 
\end{align}
is known as canonical EMT which does not have gauge field at the operator level. Therefore, the canonical OAM and SOC depends on explicit gauge choice as discussed in Jaffe–Manohar decomposition~\cite{Jaffe:1989jz}. 

Interestingly, the canonical SOC can be also represented in terms of moments of certain Generalized Transverse Momentum distribution(GTMD) or the phase-space Wigner distribution \cite{Lorce:2011kd,Lorce:2013bja,Meissner:2008ay}. Through the Fourier transform, we can relate the Wigner distributions to the Generalized Transverse Momentum distribution(GTMDs). The GTMDs are related to the Wigner correlator through the following expression  \cite{Meissner:2008ay,Tan:2021osk}:
\begin{align}
	W^{[\Gamma]}(x,\boldsymbol{k}_\perp,\Delta;n) 
	&= \int \frac{dz^-\,d^2\boldsymbol{z}_\perp}{2(2\pi)^3}\,
	e^{i x P^+ z^- - i \boldsymbol{k}_\perp\cdot \boldsymbol{z}_\perp} 
	\langle p'|\bar\psi(-\tfrac{z}{2})\,\Gamma\,\mathcal{W}\,\psi(\tfrac{z}{2})|p\rangle
	\Big|_{z^+=0},
\end{align}
where $\mathcal{W}$ denotes the gauge link ensuring color gauge invariance which is unity since we are working in light cone gauge here. 
For a spinless target, there are four twist–2 GTMDs: $F_{1,1}$, $G_{1,1}$, $ H_{1}^{\Delta}$ and $H_{1}^{k}$.
\begin{align}
W^{[\gamma^{+}]} &= F_{1},
\tag{3.7}
\\[6pt]
W^{[\gamma^{+}\gamma_{5}]} &=
\frac{ i \varepsilon_{T}^{ij}\,k_{T}^{\, i} \Delta_{T}^{\, j} }{M^2}\,
G_{1,1},
\tag{3.8}
\\[6pt]
W^{[i\sigma^{j+}\gamma_{5}]} &=
\frac{ i \varepsilon_{T}^{ij}\, k_{T}^{\, i}}{M}\,
H_{1}^{k}
+
\frac{ i \varepsilon_{T}^{ij}\, \Delta_{T}^{\, i}}{M}\,
H_{1}^{\Delta}.
\end{align}

The projection with $\Gamma=\gamma^+\gamma_5$ isolates $G_{1,1}$, which encodes the correlation between the quark spin and its orbital motion. Its $k_T^2$–moment defines the canonical SOC ~\cite{Lorce:2011kd, Lorce:2014mxa}:
\begin{align}
	\hat{C}^q_{z} &= \int dx\, d^2k_T \, \frac{k_T^2}{M^2} \,
	G_{1,1}(x,k_T^2,\Delta=0).
	\label{eq:SOC-GTMD}
\end{align}

\section{Light-front wave function for the pion }


We adopt a light-front Hamiltonian framework ~\cite{Brodsky:1997de} to describe the internal structure of the pion in terms of its quark and gluon degrees of freedom. Light-front dynamics offers a natural Hamiltonian formulation of QCD, where hadronic states are expanded in terms of{{ multiparton light-front wave functions (LFWFs) in Fock space, which are Boost invariant~\cite{Brodsky:1989pv, Harindranath:1996hq,Choi:1996mq}.}} This formulation allows for a consistent treatment of relativistic effects.   The leading Fock sector of the pion consists of a valence quark-antiquark pair. Beyond the valence sector, the pion state receives contributions from higher Fock states involving additional gluons and sea quarks. {{The Fock state expansion for the pion can be written as}} \cite{Harindranath:1998pd}: 
\be
|\pi\rangle = \sqrt{\mathcal{N}} \Big( 
\sum_{\sigma_1, \sigma_2} \psi_2^{\sigma_1 \sigma_2} |q\bar{q}\rangle 
+ \sum_{\sigma_1, \sigma_2,\lambda_3} \psi_3^{\sigma_1 \sigma_2 \lambda_3} |q\bar{q}g\rangle 
+ \cdots \Big),
\label{pionstate}
\ee
where $\mathcal{N}$ ensures the proper normalization of the full state and $\sigma_1, \sigma_2, \lambda_3 $ denote the quark, antiquark and gluon helicities. Here, $\psi_2^{\sigma_1 \sigma_2}$ and $\psi_3^{\sigma_1 \sigma_2 \lambda_3} $ denote the light-front wave functions (LFWFs) of the two- and three-particle Fock sectors, respectively. The function $\psi_2$ represents the probability amplitude for finding a quark-antiquark pair within the pion, while $\psi_3$ describes the amplitude for finding a quark-antiquark pair accompanied by a gluon.  LFWFs parameterize the complete quantum state of a hadron in terms of its partonic constituents and encodes their momentum, spin and orbital motion information.

The two-particle light-front wave function (LFWF) $\psi_2^{\sigma_1 \sigma_2}$ in the above Eq. (\ref{pionstate}) is factorized into a {spin-dependent part} and a {momentum-dependent part}. The spin part of the wave function is obtained by using the SU(6) instant-form quark model
\begin{align}
    \chi = \frac{\chi_1^\uparrow \chi_2^\downarrow-\chi_2^\uparrow \chi_1^\downarrow}{\sqrt{2}} ,
\end{align}
where the LHS refers to the pion spin state, and the RHS is a combination of the spin state of a quark and an antiquark. These spinors are transformed to the light front spin spinors through the Melosh–Wigner rotation ~\cite{Tan:2021osk,Brodsky:2000ii,Ma:1991xq,Xiao:2003wf}.
\begin{align}
    \chi_{i_{LF}}^\uparrow &= \omega_i[(q_i^+ +m_i)\chi_i^\uparrow-q_i^R\chi_i^\downarrow], \\
    \chi_{i_{LF}}^\downarrow &= \omega_i[(q_i^+ +m_i)\chi_i^\downarrow-q_i^R\chi_i^\uparrow],
\end{align}
where $ \omega^i=\frac{1}{\sqrt{2q_i^+(q_i^0+m_i)}}, q_i^+=q_i^0+q_i^3 \text{ and } q_i^{R,L}=q_i^1\pm iq_i^2$ for $i^{th}$ parton. Using $m_1=m_2=m, q_1^\mu = (k^0,\boldsymbol{k}) \text{ and } q^\mu_2=(k^0,-\boldsymbol{k})$, the spin part of the leading Fock state pion wavefunction in the light-front form is simplified to: 

\be
\begin{aligned}
	\psi_2^{\uparrow,\downarrow}(x,\mathbf{k}_\perp) &= +\frac{m}{\sqrt{m^2 + \mathbf{k}_\perp^2}} \varphi_\pi, & [l^z=0,J_z=0] \\[2mm]
	\psi_2^{\downarrow,\uparrow}(x,\mathbf{k}_\perp) &= -\frac{m}{\sqrt{m^2 + \mathbf{k}_\perp^2}} \varphi_\pi, & [l^z=0,J_z=0] \\[2mm]
	\psi_2^{\uparrow,\uparrow}(x,\mathbf{k}_\perp) &= -\frac{k^1 - i k^2}{\sqrt{m^2 + \mathbf{k}_\perp^2}} \varphi_\pi, & [l^z=-1,J_z=0] \\[2mm]
	\psi_2^{\downarrow,\downarrow}(x,\mathbf{k}_\perp) &= -\frac{k^1 + i k^2}{\sqrt{m^2 + \mathbf{k}_\perp^2}} \varphi_\pi, & [l^z=+1, J_z=0] \\[2mm]
\end{aligned}
\label{spinLFWFs}
\ee
where $x$ is the longitudinal momentum fraction carried by the quark, $\mathbf{k}_\perp$ is its transverse momentum, $m$ denotes the quark mass, and $l^z$ is the projection of the orbital angular momentum along the $z$-axis. This approach has been widely used to incorporate relativistic spin effects and spin–orbit correlations of the partonic constituents, providing a realistic description of the pion’s internal spin structure \cite{Xiao:2003wf, Ma:1991xq, Tan:2021osk, Acharyya:2024enp}. The momentum-space wave function $\varphi_\pi(x,\mathbf{k}_\perp)$ is inspired by the {AdS/QCD correspondence}, which provides a semiclassical approximation of the pion wave function \cite{Brodsky:2007hb,Bacchetta:2017vzh,deTeramond:2008ht,Brodsky:2003px,Choudhary:2023unw}:  
\be
\varphi_\pi(x,\mathbf{k}_\perp) = \frac{4 \pi}{\kappa \sqrt{x(1-x)}} 
\exp\Biggl[-\frac{\mathbf{k}_\perp^2 + m^2}{\kappa^2 \, x^\alpha (1-x)^\beta}\Biggr],
\label{scalarLFWFs}
\ee
where $\alpha, \beta$ control the shape of the longitudinal momentum distribution in $x \rightarrow 0 $ and $ x\rightarrow 1$ limits respectively. The parameter $\kappa$ is known as AdS/QCD scale parameter and it is related to the confinement scale~\cite{Brodsky:2008pg,deTeramond:2008ht,Ahmady:2018muv}.

The pion state can be extended beyond the valence two-particle state by adding gluons and sea quarks \cite{Brodsky:2007hb,Pasquini:2023aaf, Lan:2025qio,Lan:2024ais}. However, going beyond the leading Fock sector is theoretically and computationally demanding, as the treatment of three- or many-body systems on the light front is nontrivial.  
In this work, we model the next-to-leading Fock sector LFWF for the pion by using the analytic form of the three-particle (quark-antiquark-gluon) LFWF of a state obtained using perturbative light-front QCD Hamiltonian \cite{Harindranath:1998pd}.  The three-particle LFWF thus incorporates  the  effect of the quark–antiquark–gluon $ q\bar{q}g $ interactions. In the light-front framework, this amplitude arises from the quark–gluon–quark interaction terms in the Hamiltonian, which also contains the relevant gluon self-interaction contributions at leading order \cite{Harindranath:1998pd}. The inclusion of a gluon in the pion state allows the estimate of the gluon contribution to the properties of the pion, including the spin-orbit correlation. We express the three-particle LFWFs $ \psi_3^{\sigma_1\sigma_2\lambda_3} $  using the light-front eigenvalue equation in terms of the two-particle LFWF and the quark-antiquark-gluon vertex using the framework discussed in  \cite{Harindranath:1998pd,Harindranath:1996sj,Harindranath:1998pc}

\begin{eqnarray}
	\psi_3^{\sigma_1 \sigma_2 \lambda_3}(x_1, k_1^{\perp}; x_2, k_2^{\perp};
	1-x_1-x_2, k_3^{\perp}) = {\cal M}_1 + {\cal M}_2,
\end{eqnarray}
with
\begin{eqnarray}
	{\cal M}_1 = && { 1 \over E} (-) { g \over \sqrt{2 (2 \pi)^3}} T^a
	{ 1 \over \sqrt{1 - x_1 - x_2}} ~V_1~ 
	\psi_2^{\sigma_1' \sigma_2}(1-x_2, -k_2^\perp; x_2,k_2^\perp)
\end{eqnarray}
and
\begin{eqnarray}
	{\cal M}_2 = && { 1 \over E} { g \over \sqrt{2 (2 \pi)^3}} T^a
	{ 1 \over \sqrt{1 - x_1 - x_2}} ~V_2~
	\psi_2^{\sigma_1 \sigma_2'}(x_1,k_1^\perp;1-x_1,-k_1^\perp)
\end{eqnarray}
where
\begin{eqnarray}
	V_1=\chi_{\sigma_1}^\dagger \sum_{\sigma_1'}
	\big [ { 2 k_3^\perp \over 1 - x_1 -x_2} - { (\sigma^\perp. k_1^\perp
		- i m) \over x_1} \sigma^\perp + \sigma^\perp {(\sigma^\perp. k_2^\perp -
		im) \over 1-x_2} \big] \chi_{\sigma_1'}. (\epsilon^\perp_{\lambda_1})^*,
\end{eqnarray}
and
\begin{eqnarray}
	V_2=\chi_{-\sigma_2}^\dagger \sum_{\sigma_2'}
	\big [ { 2 k_3^\perp \over 1 - x_1 -x_2} - \sigma^\perp
	{ (\sigma^\perp. k_2^\perp
		- i m) \over x_2}  +  {(\sigma^\perp. k_1^\perp -
		im) \over 1-x_1} \sigma^\perp 
	\big] \chi_{-\sigma_2'}. (\epsilon^\perp_{\lambda_1})^*.
\end{eqnarray}
Here $x_1, x_2$ are the longitudinal momentum fractions of quark and anti-quark and the gluon takes remaining $1-x_1-x_2$. The {{relative}}  transverse momentum of the quark, anti quark and gluon are $k_1^{\perp}, k_2^{\perp}$ and $k_3^{\perp}=-k_1^{\perp}-k_2^{\perp}$ respectively. The spin states of quark and antiquark are encoded in $\chi_{\sigma_i}$ whereas $\epsilon^\perp_{\lambda_1}$ denote the polarization state of gluon.
$ {\cal M}_1$ and $ {\cal M}_2$ correspond to two possible ways of emitting a gluon either from the quark or from the anti-quark starting from the pion two particle state $ q \bar{q}$. Each term contains the quark–gluon coupling constant, color factor $T_a$, vertex structures $V_1$, $ V_2$, and the energy denominator E, explicit expressions of which are given in the Appendix \ref{3lfwf} and ~\cite{Harindranath:1998pd}. The three particle wave functions $\psi_3^{\sigma_1 \sigma_2 \lambda_3}$ can also be viewed as combinations of spin- and momentum-dependent components and they also depend on the two-particle amplitudes $\psi_2^{\sigma_1 \sigma_2}$.

 {{ Thus in this formalism, one can obtain an analytic form of  the three-particle wave function  once the two-particle LFWFs are specified. In this work, we use the model LFWFs in the two-particle sector as given in Eq. (\ref{spinLFWFs}) and (\ref{scalarLFWFs}).}}  Different combinations of quark, antiquark, and gluon helicities correspond to states with distinct total orbital angular momentum $l_z$. For instance, the states $\psi_3^{\uparrow \uparrow -1}$ and $\psi_3^{\downarrow \downarrow +1}$ correspond to $l_z = 0$, while configurations such as $\psi_3^{\uparrow \downarrow -1}$, $\psi_3^{\downarrow \uparrow +1}$, $\psi_3^{\downarrow \uparrow -1}$, and $\psi_3^{\uparrow \downarrow +1}$ correspond to $l_z = \pm 1$. The states $\psi_3^{\uparrow \uparrow +1}$ and $\psi_3^{\downarrow \downarrow -1}$ correspond to $l_z = \pm 2$.  We are considering the LFWFs upto $l_z = \pm 1 $ as in two particle case Eq. (\ref{spinLFWFs}) and (\ref{scalarLFWFs}). The detailed expressions of three particle LFWFs are given in Appendix \ref{3lfwf}. An interesting feature of the three-particle LFWFs is that their orbital angular momentum structure arises from a combination of different two-particle components. Even the three-particle states with $l_z = 0$ depend not only on the $l_z = 0$ two-particle wave functions but also on the $l_z = 1$ components. Likewise, the three-particle states with $L_z = \pm 1$ receive contributions from both $l_z = 0$ and $l_z = 1$ two-particle components. This mixing reflects the nontrivial effects of {{quark-gluon interaction.}}  Once the two-particle amplitudes are fixed, the analytic forms of all corresponding three-particle LFWFs can be systematically obtained within this framework.

{{The next step is to construct the parton distribution functions (PDFs) using the light-front overlap representation. In particular, the unpolarized quark distribution $f_1^q(x)$ can be expressed in terms of contributions from both two-particle and three-particle LFWFs as follows:}}
\be
f_1^q(x) = \frac{1}{16\pi^3}\int d^2\mathbf{k}_\perp \sum_{\sigma_1,\sigma_2} 
\left|\psi_2^{\sigma_1 \sigma_2}(x, \mathbf{k}_\perp)\right|^2
+ \frac{1}{16\pi^3} \int dx_2\, d^2\mathbf{k}_\perp\, d^2\mathbf{k}_{2\perp} 
\sum_{\sigma_1,\sigma_2,\lambda_3} 
\left|\psi_3^{\sigma_1 \sigma_2 \lambda_3}(x, \mathbf{k}_\perp; x_2, \mathbf{k}_{2\perp})\right|^2 .
\label{unpolPDF}
\ee
In the model that we use, the light-front wave functions introduced above depend on a set of parameters that characterize the internal dynamics of the pion, namely the constituent quark mass $m$, and the parameters $\alpha$, $\beta$, and $\kappa$ controlling the shape and scale of the momentum distribution. These parameters enter both the analytic forms of the LFWFs and the resulting parton distribution functions (PDFs). These parameters must be determined by comparison with phenomenological data to fully specify and constrain the model. 
We perform a fit of the calculated $f_1^q(x)$ to the global QCD analysis results of the JAM Collaboration for the negatively charged pion at next-to-leading order (NLO) {which is one of the most recent phenomenological extractions of mesons} ~\cite{Barry:2021osv}. We focus on the $\bar{u}$ quark distribution in the $\pi^-$ within the kinematic range $0.001 < x < 1$ at the input scale $Q=1~\mathrm{GeV}$, sampled at 100 uniformly spaced points from the JAM21 NLO set. 
The model prediction is evaluated at the same kinematic points, and the best-fit parameters are obtained through a standard $\chi^2$ minimization procedure. 
During the fit, the pion mass is fixed to $M_\pi = 0.134~\mathrm{GeV}$ and the quark mass is fixed to $m_q = 0.003~\mathrm{GeV}$. The free parameters of the model include the normalization $N$, the longitudinal shape parameters $\alpha$ and $\beta$, and the transverse confinement scale $\kappa$, and the quark–gluon coupling $g$. 
The parameters $\alpha$ and $\beta$ control the asymptotic behavior of the parton distributions in the limits $x \to 0$ and $x \to 1$, respectively~\cite{Brodsky:1989db,Brodsky:1994kg}. 
The confinement scale typically lies in the range $\kappa \approx 0.2$–$0.6~\mathrm{GeV}$, providing good fits to pion electromagnetic form factors as also found in previous pion analyses ~\cite{Bacchetta:2017vzh}. 
However, once gluonic degrees of freedom are explicitly incorporated in the pion’s light-front wave function, our fit indicates a significantly higher confinement scale $\kappa = 2.30^{+0.06}_{-0.43}~\mathrm{GeV}$. A similar upward shift in $\kappa$ has been observed in gluon–spectator models of the proton~\cite{Sain:2025kup} which suggests that gluon dynamics effectively enhance the transverse momentum width of the pion’s internal structure. 
The resulting parameter uncertainties are estimated from the covariance matrix of the fit within $1\sigma$ confidence level. 
The obtained $\chi^2$ per degree of freedom ($\chi^2/\mathrm{dof} \approx 0.6$) demonstrates an excellent agreement between the model prediction and the JAM21 NLO data across the entire $x$ range. The model parameters are summarized in Table~\ref{tab:parameter}.
\begin{figure}[h]
    \centering
    \includegraphics[width=0.4\linewidth]{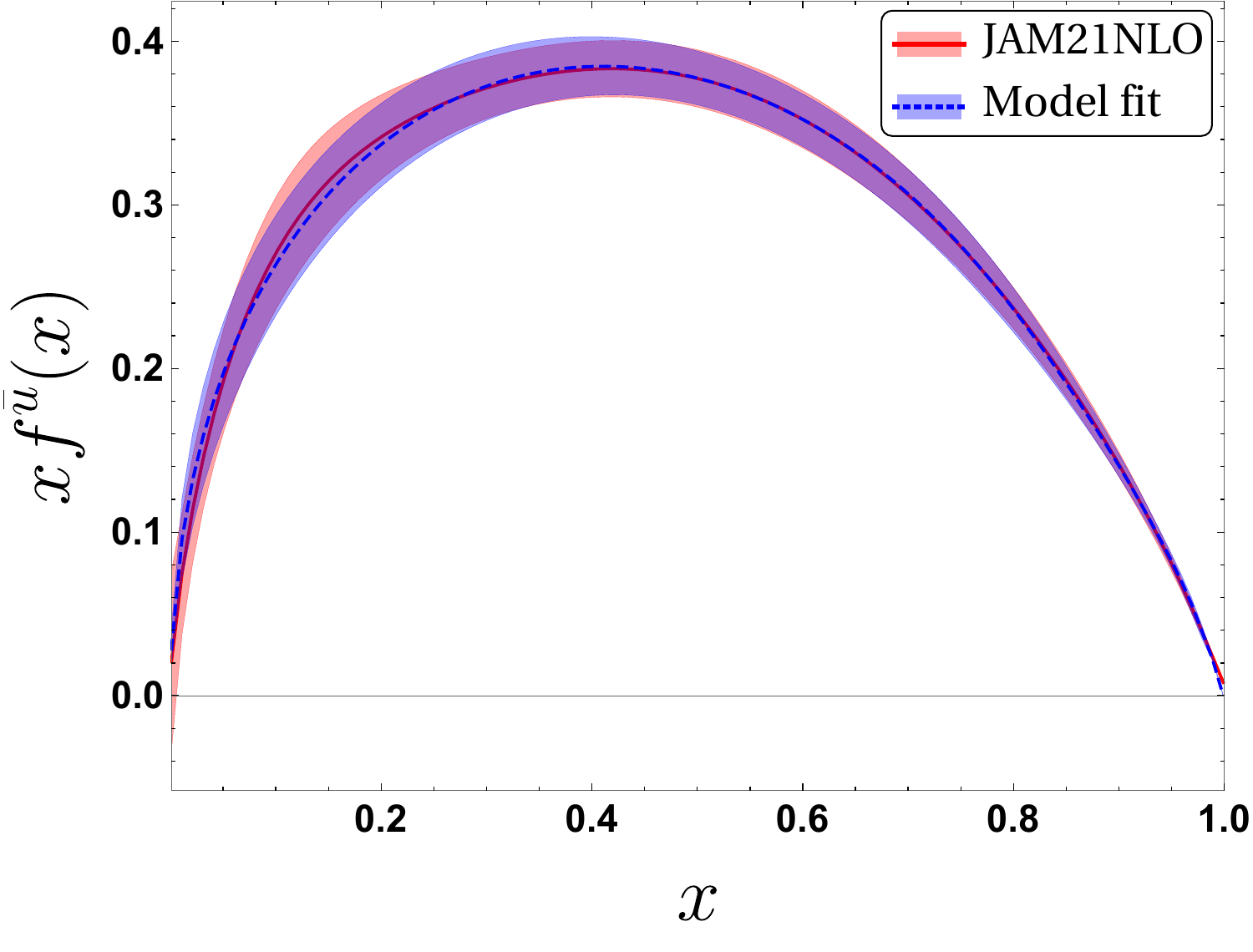}
    \includegraphics[width=0.4\linewidth]{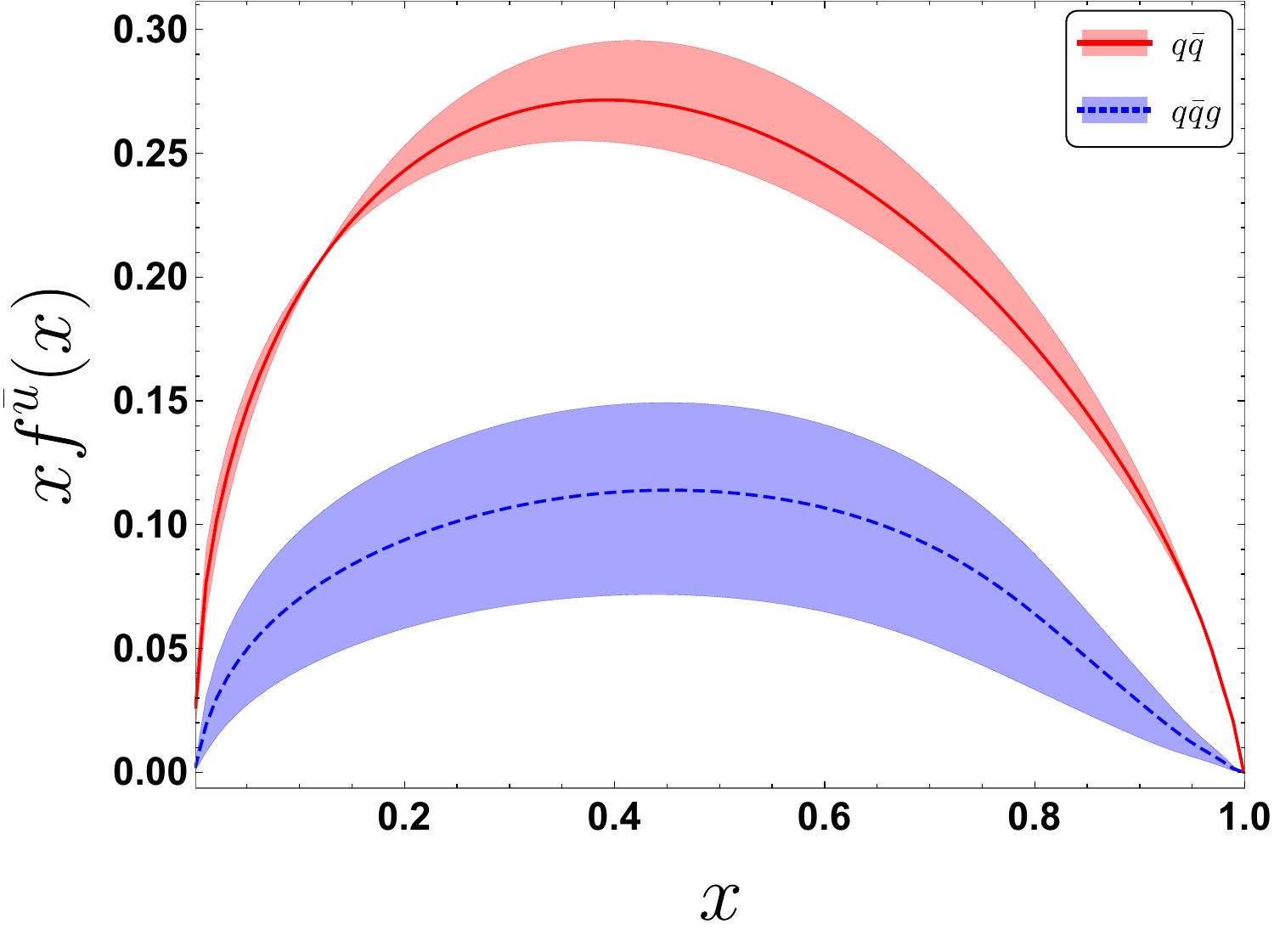}
    \caption{Left: $\pi^-$ antiquark distribution $x f^{\bar{u}}(x)$ as a function of $x$ at $Q_0 = 1~\mathrm{GeV}$, compared with JAM21 NLO results~\cite{Barry:2021osv}. Right: Separate contributions from the $|q\bar{q}\rangle$ and $|q\bar{q}g\rangle$ Fock sectors illustrating their relative impact.}
    \label{pdfplot}
\end{figure}

\begin{table}[h] 
	\centering
	\renewcommand{\arraystretch}{1.3} 
	\setlength{\tabcolsep}{10pt} 
	\begin{tabular}{|c|c|c|c|c|c|}
		\hline\hline
		Parameter & $N$ & $\alpha$ & $\beta$ & $\kappa$ & $g$\\
		\hline
        Values with $1\sigma$ uncertainties
    & $0.765^{+0.074}_{-0.055}$
    & $0.449^{+0.083}_{-0.069}$
    & $1.697^{+0.047}_{-0.041}$
    & $2.297^{+0.060}_{-0.425}$
    & $0.939^{+0.188}_{-0.251}$ \\
		\hline\hline
	\end{tabular}
    \caption{Parameters with uncertainities. {\PC }}
    \label{tab:parameter}
\end{table}

The left panel of Fig.~\ref{pdfplot} illustrates the quality of the fit between our model predictions and the JAM21 NLO PDF~\cite{Barry:2021osv}. The total unpolarized quark PDF obtained by combining the $|q\bar{q}\rangle$ and $|q\bar{q}g\rangle$ contributions fit reasonably well  with the JAM21 analysis in $0.001<x<1$ range. The right panel displays the separate unpolarized quark distributions from the two Fock sectors, indicating that the two-particle ($|q\bar{q}\rangle$) component dominates over the three-particle ($|q\bar{q}g\rangle$) one. The simultaneous fit of both Fock components with similar parameter values yields a consistent and stable description of the data, with the $|q\bar{q}\rangle$ sector exhibiting smaller uncertainties compared to $|q\bar{q}g\rangle$ at low $x$. In particular, the uncertainty band remains narrow at small $x$ and becomes nearly a line around $x \approx 0.12$ for the leading Fock sector. The apparent vanishing of the error band in the two-particle model $x f^{\bar{u}}(x)$ occurs because the parameters were obtained by fitting the combined contributions from both the two- and three-particle Fock sectors to the data.
 
\section{Numerical results and analysis}
In this section, we present the numerical results of GPDs, GTMDs and spin-orbit correlation, discuss the advantages of going beyond the leading Fock sector for pion, and show the effect of dynamical interactions with gluons. For the numerical analysis, we have used the parameters and their uncertainties as shown in the Table. \ref{tab:parameter}.

\begin{figure}[ht]
		\begin{minipage}{0.49\linewidth}
			\includegraphics[scale=1.7
            ]{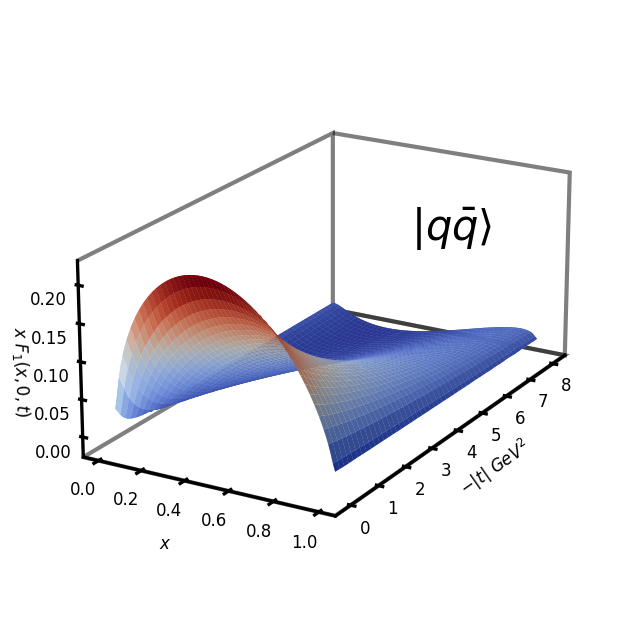}
		\end{minipage}
		\begin{minipage}{0.49\linewidth}
			\includegraphics[scale=1.7]{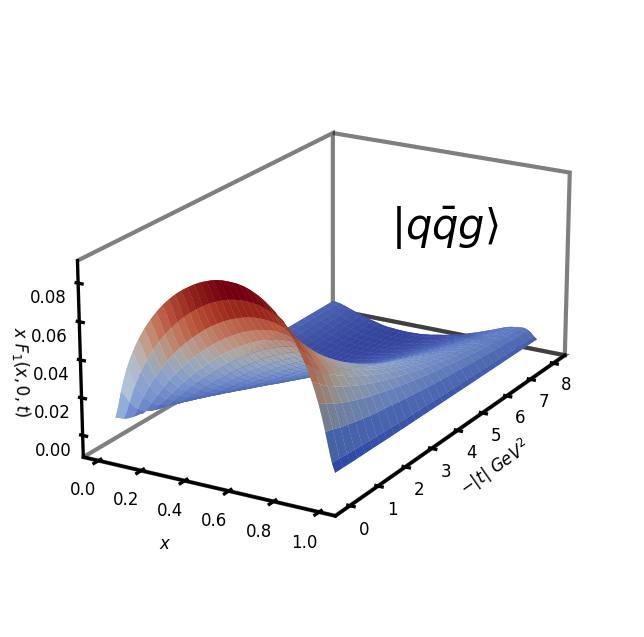}
		\end{minipage}  
		\caption{The left and right plot show results for the unpolarized chiral-even quark GPD for $|q\bar{q} \rangle$ and $|q\bar{q} g \rangle$, respectively, as a function of $x$ and $-|t|$.}
		\label{gpd1}
	\end{figure}

In Fig.~(\ref{gpd1}), the unpolarized chiral-even quark GPD is shown as a function of the longitudinal momentum fraction $x $ and momentum transfer squared $t$ for both the $|q\bar{q} \rangle$ and $|q\bar{q}g \rangle$  Fock sectors. As $t$ increases, the peak of the distribution shifts towards higher $x$  which is also consistent with the behavior of the collinear PDF (corresponds to GPD at $t=0 $) at large $x$. A similar trend is observed in both Fock sectors; however, the contribution from the higher Fock component $|q\bar{q} g \rangle$  is significantly smaller than that from the leading $|q\bar{q} \rangle$  sector. The shift of the peak to higher $x$ with increasing $-|t|$ is more pronounced in the $|q\bar{q}g \rangle$ sector. The chiral-odd GPD contributions are found to be negligible compared to the unpolarized chiral-even GPD in both Fock sectors.

\begin{figure*}[htp]

\includegraphics[width=0.46\textwidth]{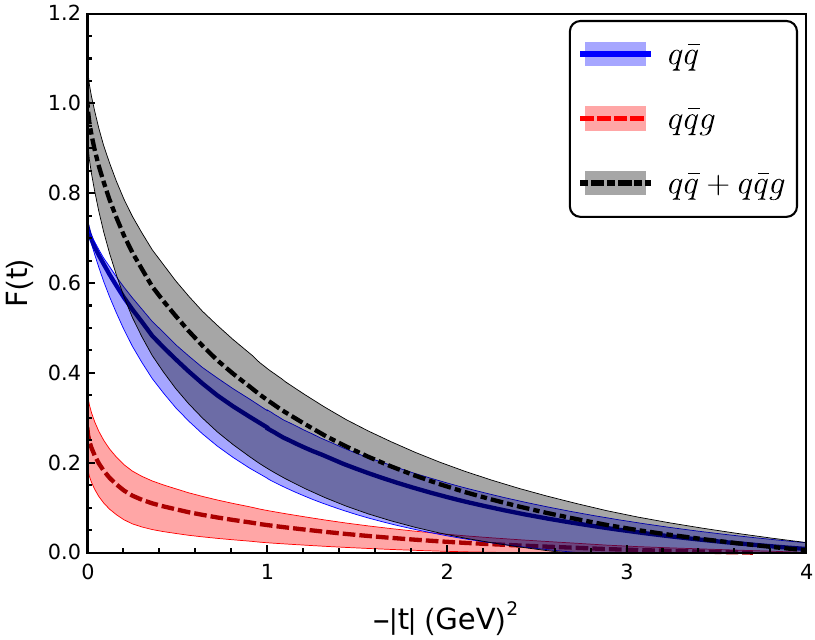}\quad\quad
\includegraphics[width=0.46\textwidth]{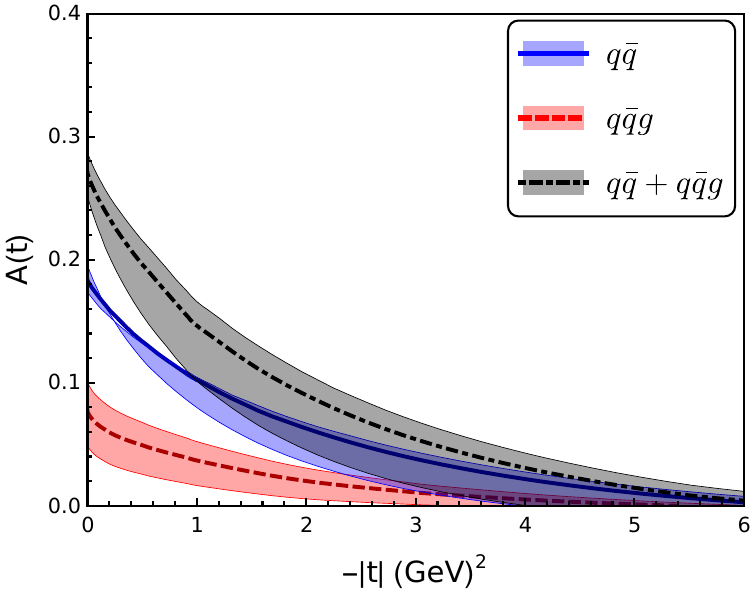}\\
 \caption{The left and right plot display the Dirac form factor, $F(t)$ and the gravitational form factor, $A(t)$ respectively as a function of momentum transfer squared, $-|t|$ .}
 \label{FFs}
\end{figure*}

The left plot in Fig.~\ref{FFs} shows the contributions of the $|q\bar{q} \rangle$ and $|q\bar{q} g\rangle$ Fock sectors to the first moment of the GPD $F(t)$, which corresponds to the Dirac form factor of the pion. The value $F(0) \approx 1$ reflects the charge sum rule, which is correctly satisfied in our model when both the valence and higher Fock components are included. The $|q\bar{q} g\rangle$ contributes around $25 \%$ of total charge which reflects that even though gluon has no charge but the presence of quark-gluon interaction provides significant contribution to the charge sum rule. 
{The bands in Fig.~\ref{FFs} represent the uncertainty in the values of fitted parameters as shown in Table.~\ref{tab:parameter}.}
The $ F^{q \bar{q} g}(t)$ falls off more rapidly than $ F^{q \bar{q} }(t)$; in fact, it vanishes around $ -|t|=2 $ GeV$^2$, whereas the $q \bar{q}$ component remains significant up to $ -|t|=4 $ GeV$^2$. This means that the quark–antiquark–gluon configuration contributes predominantly at larger distance scales (lower momentum transfers), thus highlighting the role of gluonic degrees of freedom in shaping the pion's internal structure and charge distribution. 
\begin{figure}[ht]
\centering
    \begin{minipage}{0.49\linewidth}
			\includegraphics[scale=0.55]{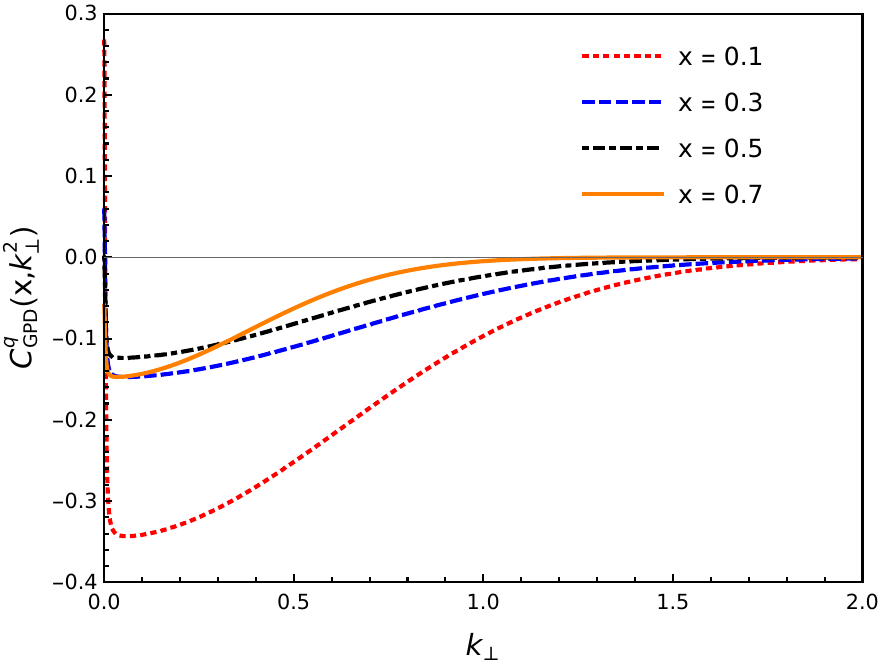}
	\end{minipage}
	\begin{minipage}{0.49\linewidth}
			\includegraphics[scale=0.63]{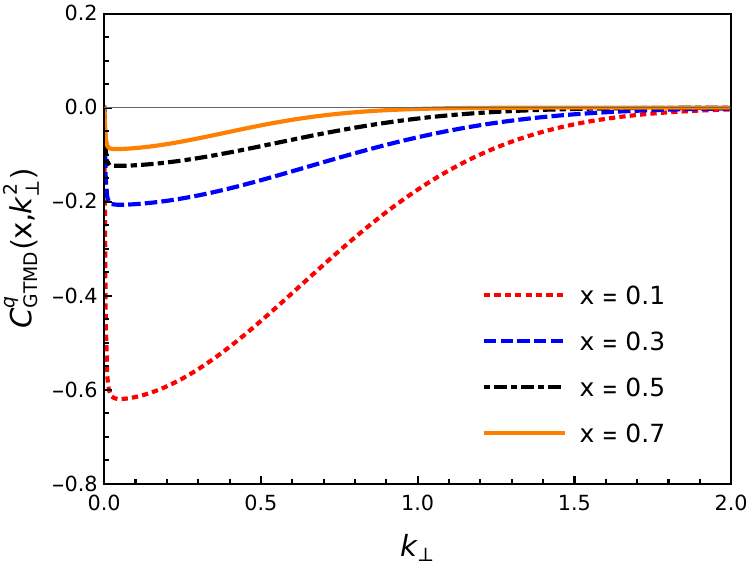}
	\end{minipage}  
    \caption{The left and right plot represent the $C^q_z$ obtained from GPD and GTMD, respectively, plotted as a function of transverse momentum $k_\perp$ at fixed values of longitudinal momentum fraction $x$. Both the plots shows contributions from $|q\bar{q}\rangle$ Fock sector only.}
    \label{Cxk}
\end{figure}

{{The right plot in Fig.~\ref{FFs} presents the contributions of the same Fock sectors  to the gravitational form factor $A(t)$.}} The value $A(0)$ represents the fraction of the pion's momentum carried by the quark, in which the $ q \bar{q} g$ Fock sector contributes significantly around $30-40 \%$ of total mass from quark.
Both $F(t)$ and $A(t)$ decrease with increasing $-|t|$. This behavior reflects the fact that higher momentum transfers probe the pion at shorter distance scales. Larger $|t|$ corresponds to resolving finer and more localized structures within the pion. 
The fall-off of $F(t)$ indicates that the pion's charge is distributed over a finite spatial region, whereas the decrease of $A(t)$ shows that the momentum carried by the constituents is also similarly spread. In both cases, the dominant contributions arise from the valence $|q\bar{q}\rangle$ sector, while the $|q\bar{q}g\rangle$ sector introduces small corrections, capturing the effects of gluonic degrees of freedom and providing a more complete picture of the pion's internal structure.

\begin{figure}[ht]
\centering
    \begin{minipage}{0.49\linewidth}
			\includegraphics[scale=0.65]{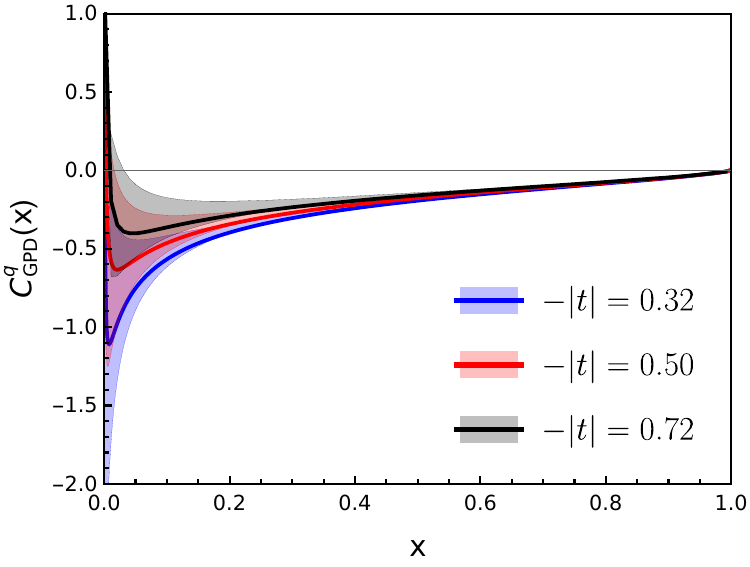}
	\end{minipage}
	\begin{minipage}{0.49\linewidth}
			\includegraphics[scale=0.65]{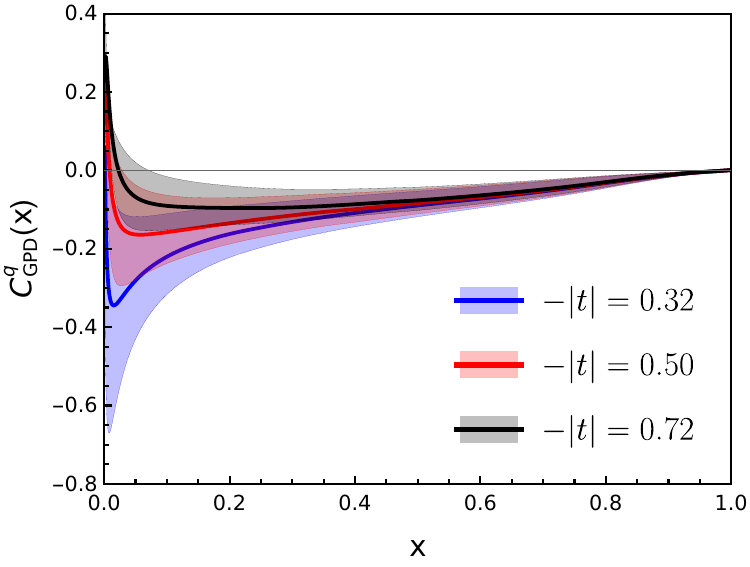}
	\end{minipage}  
    \caption{The left and right plots represent the contribution of $|q\bar{q}\rangle$ and $|q\bar{q}g\rangle$ to the $C^q_{z, kin}$ plotted as a function of longitudinal momentum fraction $x$ at fixed values of momentum transfer squared $-|t|$.}
    \label{CxGPD23}
\end{figure}

\begin{figure}[ht]
\centering
    \begin{minipage}{0.49\linewidth}
			\includegraphics[scale=0.65]{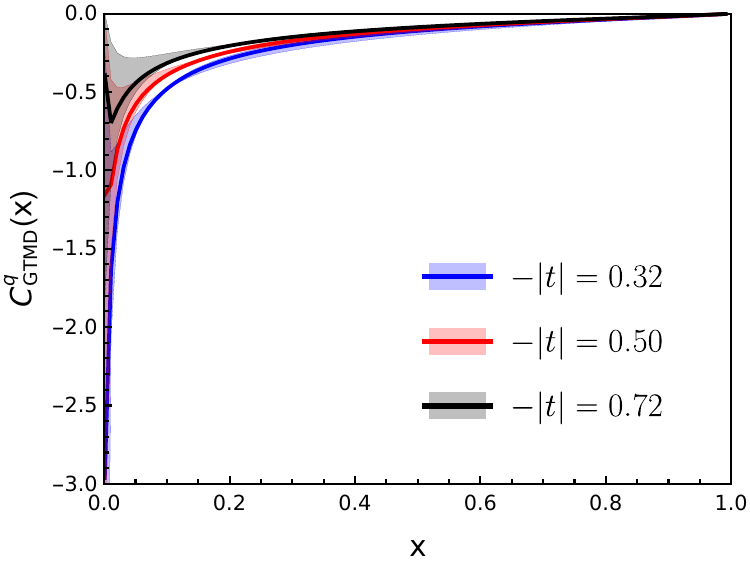}
	\end{minipage}
	\begin{minipage}{0.49\linewidth}
			\includegraphics[scale=0.49]{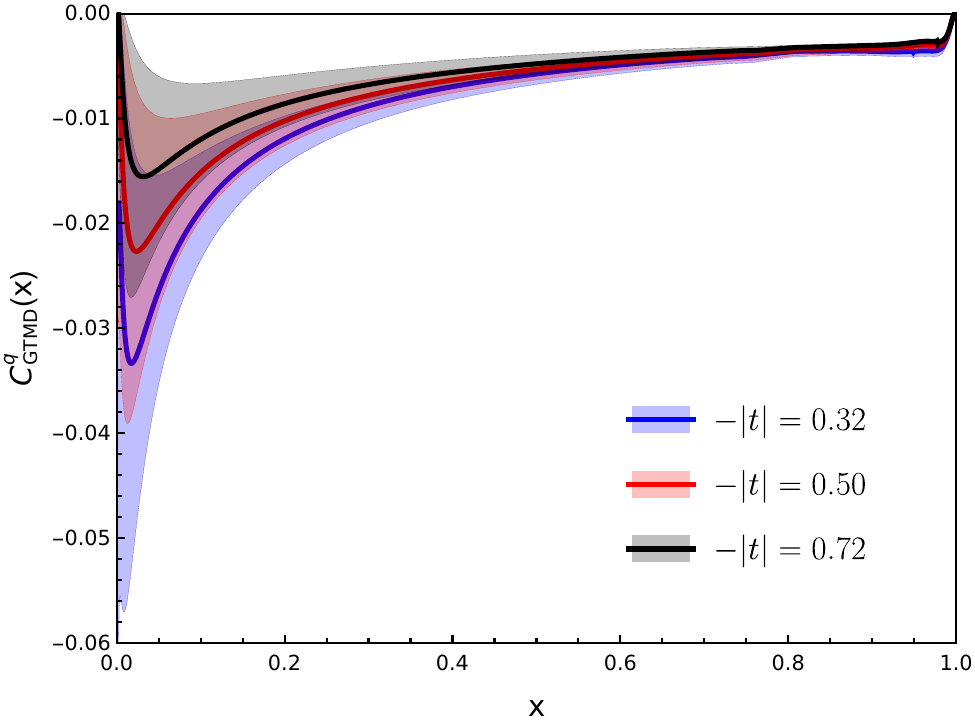}
	\end{minipage}  
    \caption{The left and right plots represent the contribution of $|q\bar{q}\rangle$ and $|q\bar{q}g\rangle$ to the $C^q_{z}$ plotted as a function of longitudinal momentum fraction $x$ at fixed values of momentum transfer squared $-|t|$.}
    \label{CxGTMD23}
\end{figure}

The analysis of spin-orbit correlations offers additional insight into the interplay between quark spin and orbital angular momentum, complementing the information revealed by the form factors $F(t)$ and $A(t)$. We start this by showing the $\boldsymbol{k_\perp}$ dependence of the kinetic and canonical $C^q_z$ for fixed values of $x$, shown respectively in the left and right panels of Fig. \ref{Cxk}. The analysis includes only the contributions from the $|q\bar{q}\rangle$ Fock sector. Consistent with the observations in Ref. \cite{Tan:2021osk}, we find that $C^q_{\text{GTMD}}(x, \boldsymbol{k}^\perp)$ remains negative over the entire $\boldsymbol{k_\perp}$ range, whereas $C^q_{\text{GPD}}(x, \boldsymbol{k}^\perp)$ becomes positive at lower values of $k_\perp$. 
The transition to positive values occurs at much smaller $k_\perp$, is significantly sharper, and is observed for all values of $x$ considered here. {The reason for this difference is the quark mass multiplied to the chiral odd GPD in Eq. \eqref{CqGPD}. Since we take a much smaller quark mass than in \cite{Tan:2021osk}, the chiral odd contribution to the $C^q_{\text{GPD}}(x, \boldsymbol{k}^\perp)$ becomes relevant at very low values of $k_\perp$. Additionally, the difference in the spin-independent part of the two models is the reason behind a very steep rise in our case.} In both approaches, the correlations vanishes for $k_\perp >2 $ and all values of $x$. We have verified that the sharp rise in the $C^q_{\text{GPD}}(x, \boldsymbol{k}^\perp)$ is because of the contribution from the chiral odd GPD $H_1$, and it vanishes in the chiral limit, $m=0$, as can be seen from Eq. \eqref{eom}. 

Next, we present the results of longitudinal spin–orbit correlation $C_q(x)$, which shows interesting physics. The longitudinal spin-orbit correlation can be evaluated in two different approaches as discussed in Sec. \ref{SOC}. Fig.~\ref{CxGPD23} presents the results obtained from the GPD (kinetic) approach, while Fig.~\ref{CxGTMD23} shows the corresponding results from the GTMD (canonical) approach, both plotted as functions of $x$ at different fixed values of $-|t|$. In each figure, the left panels display the results for the $|q\bar{q}\rangle$ state, whereas the right panels correspond to the $|q\bar{q}g\rangle$ state. It is understandable from the two plots that the anti-correlation decreases with increasing $x$ and $t$ in both approaches in both the sectors. Individually, the two-particle ($|q\bar{q}\rangle$) contributions from the two approaches exhibit visible differences, particularly in the small-$x$ region. In Ref.~\cite{Tan:2021osk}, it is reported that the kinetic and canonical SOC coincide at $ t=0$ when restricted to the leading Fock sector. In Appendix~\ref{twoParGPDGTMD}, we present detailed expressions for the two-particle correlations derived from both the GPD and GTMD frameworks, which explicitly demonstrate the difference between them. These results coincide only for specific choices of gauge and model parameterizations, as in Ref.~\cite{Tan:2021osk}. 
The presence of gluon in the pion state can affect the kinetic and canonical SOC results significantly. We observe that the SOC from the GPD approach is positive at small $x$ and changes sign at intermediate $x$, indicating a dynamical shift in the orbital–spin alignment. In contrast, the GTMD-based results remain negative throughout the entire $x$ range but with a smaller magnitude. This qualitative difference shows the effect of the quark–gluon interactions coming from the higher Fock sector in the two formulations and suggests that gluonic degrees of freedom play a crucial role in both the kinetic and canonical pictures of partonic spin–orbit dynamics. 

\begin{figure*}[htp]
\includegraphics[width=0.46\textwidth]{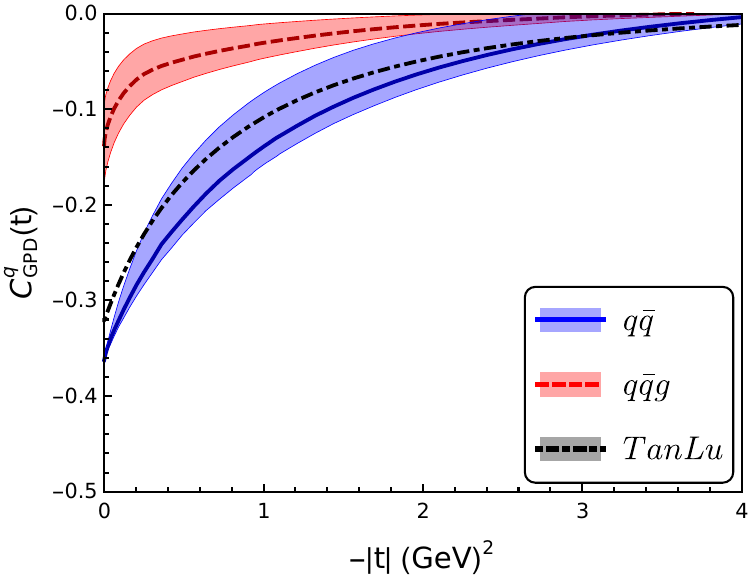}\quad\quad
\includegraphics[width=0.46\textwidth]{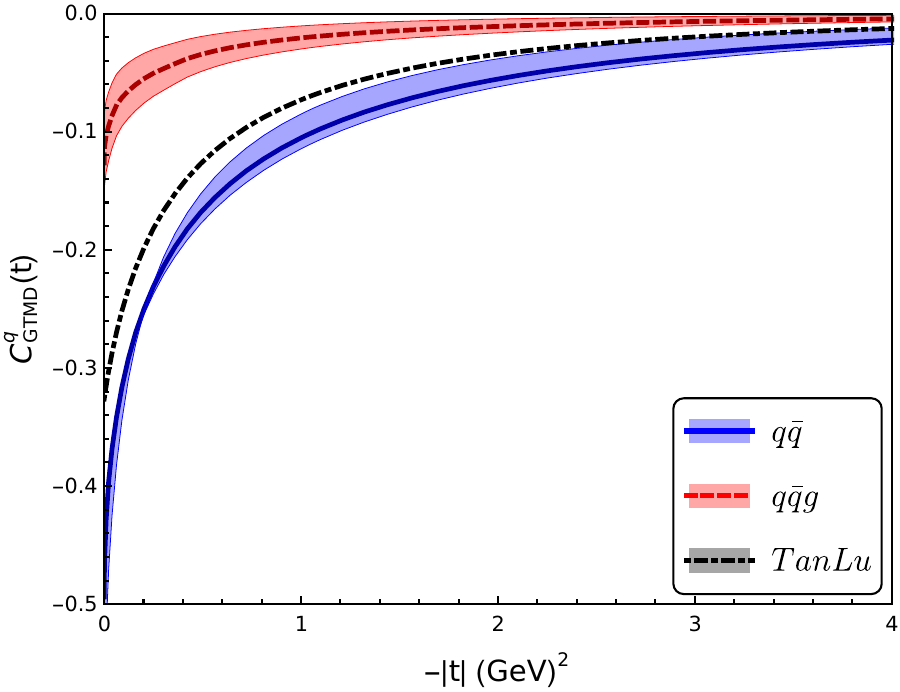}\\
 \caption{The left and right plot represent the longitudinal spin-orbit correlation $C^q_z$ obtained from GPD and GTMD, respectively, plotted as a function of momentum transfer squared, $-|t|$.}\label{Ct}
\end{figure*}

{The variation of spin-orbit correlations as a function of momentum transfer squared is shown in Fig. \ref{Ct}}. The left plot shows the $C_q(t)$ obtained from Eq. \eqref{eom} using moments of GPD, and the right one is obtained from the non-forward limit of Eq. \eqref{eq:SOC-GTMD} (We have considered $k_\perp.\Delta_\perp=0$ to keep things simple in case of GTMD). 
A comparison has been shown between the $C_q(t)$ obtained from two and three-particle sectors in our model, as well as that obtained from a two-particle pion model used in Ref. \cite{Tan:2021osk}. We observe that the results for the $ q \bar{q}$ Fock sector in our model closely overlap with those from Ref. \cite{Tan:2021osk}, while the $q \bar{q} g$ correlation in our model is comparatively smaller.
Similar to the nucleon \cite{Lorce:2014mxa}, the spin and orbital angular momentum of a quark are anti-correlated in the pion as well, and this anti-correlation decreases with an increase in momentum transfer. 

It is also important to note that the numerical value of the total spin-orbit correlation obtained from GPD is $ -0.48$, and from GTMD is $-0.68$
. In the two-particle model used by Tan and Lu in \cite{Tan:2021osk}, $C^q_{z, kin}(0)=C^q_z(0) = -0.32$. It is evident that with the inclusion of quark-gluon dynamics in the pion, the spin-orbit correlation obtained from the two different approaches is not the same. {We find that at the level of three particle sector, $C^{q/q\bar{q}g}_{z,\text{kin}}(0)=-0.12 \text{ and } C^{q/q\bar{q}g}_{z}(0)=-0.16$}. 
In our model, as shown in appendix \ref{twoParGPDGTMD},  $C^{q/q\bar{q}}_{z,\text{kin}}(0) \text{ and } C^{q/q\bar{q}}_{z}(0)$ also differ from each other, thus indicating that the matching of SOC from GPD and GTMD approaches in \cite{Tan:2021osk} may just be a model-dependent artifact as mentioned by the authors.

\begin{figure}[ht]
\centering
    \begin{minipage}{0.49\linewidth}
			\includegraphics[scale=0.65]{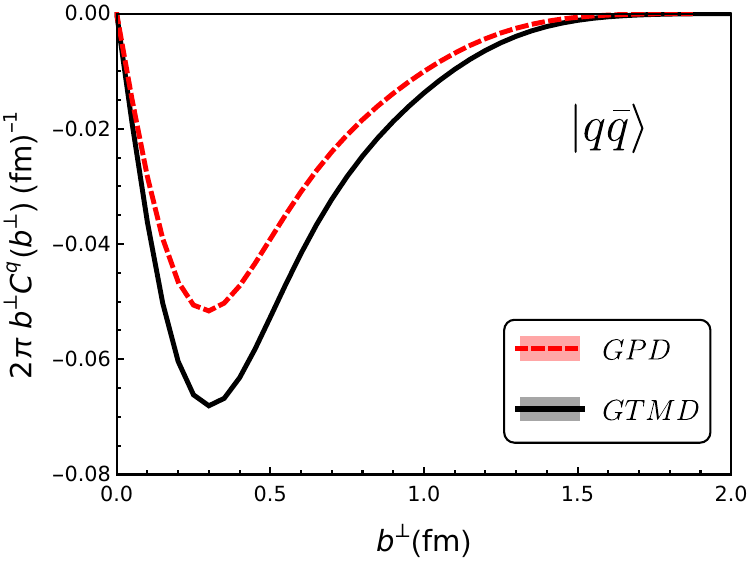}
	\end{minipage}
	\begin{minipage}{0.49\linewidth}
			\includegraphics[scale=0.65]{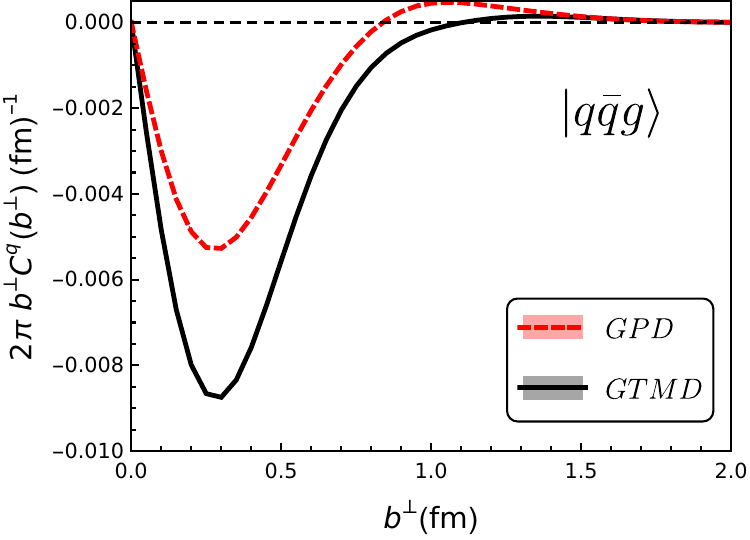}
	\end{minipage}  
    \caption{Spatial distribution of quark kinetic and canonical longitudinal SOC as a function of impact parameter $b_\perp$. Left: Contribution from $|q\bar{q} \rangle $ sector; Right: Contribution from $|q\bar{q} g\rangle $ sector}
    \label{CqSpatial}
\end{figure}

{The spatial distribution of $C^q_{z,\text{kin}}$ and $C^q_{z}$  in the impact-parameter space $b_\perp$ is shown in Fig. \ref{CqSpatial}. The left and right panels show the contributions from the $|q\bar{q} \rangle$ and $|q\bar{q} g\rangle$ sectors, respectively.}  To evaluate the spatial distribution of the spin–orbit correlation, which involves taking the Fourier transform of $C^q(t)$, we use a Gaussian wave-packet state centered at the origin \cite{Diehl:2002he, Chakrabarti:2005zm}. {The spatial distributions constructed from plane-wave states are inherently ambiguous \cite{PhysRevD.62.071503, Leader:2013jra}, thus using a localized wave packet avoids this issue, and choosing a Gaussian form for the wave packet is simpler.}
In our earlier studies, such wave-packet states were used to obtain the spatial distributions of pressure, shear forces, energy density, and angular momentum for a dressed quark state \cite{More:2021stk, More:2023pcy, Singh:2023hgu, Mukherjee:2024xnu}. {We consider the state to be localized in transverse momentum space, while carrying a definite longitudinal momentum}
\begin{align}
\frac{1}{16\pi^3}\int \frac{d^2 p^\perp dp^+}{p^+}\phi(p) \mid p^+,p^\perp,\lambda \rangle,
\end{align}
where 
$\phi(p)=p^+\ \delta(p^+-p_0^+)\ \phi\left(\boldsymbol{p}^\perp\right)$.
We choose a Gaussian form for the transverse momentum dependence,
\begin{align}
\phi\left(\boldsymbol{p}^\perp\right)
= e^{-\frac{\boldsymbol{p}^\perp}{2\sigma^2}}
\end{align}
where $\sigma$ denotes the Gaussian width which controls the localization of the state in the impact-parameter space. We choose $\sigma=1 $ GeV here; however, the choice of $\sigma$ doesn't matter much as long as $\sigma \gg 1/p^+$ \cite{Diehl:2002he}.

$C^{q/|q\bar{q}\rangle}_z(b_\perp)$ obtained from both GPD and GTMD approaches seems to show similar behavior to the one shown in Ref.~\cite{Lorce:2025ayr} whereas $C^{q/|q\bar{q}g\rangle}_z(b_\perp)$ turns ever so slightly positive towards the tail before vanishing.  We observe that $C^{q/|q\bar{q}\rangle}_z(b_\perp)$ and $C^{q/|q\bar{q}g\rangle}_z(b_\perp)$ both peak at the same point in both approaches, around $ b_\perp \approx 0.25$. The contribution of the three-particle sector to SOC is significantly smaller as compared to that of the two-particle sector. It is observed that the spatial distribution of SOC is concentrated almost entirely in the region of $ b_\perp < 1 $ fm. {The difference between $C^{q/|q\bar{q}\rangle}_z(b_\perp)$ and $C^{q/|q\bar{q}\rangle}_{z,\text{kin}}(b_\perp)$ in the left panel of Fig.  \ref{CqSpatial} can be traced to the choice of model as shown in Appendix~\ref{twoParGPDGTMD}. The right panel of Fig.  \ref{CqSpatial} shows the difference between $C^{q/|q\bar{q}g\rangle}_z(b_\perp)$ and $C^{q/|q\bar{q}g\rangle}_{z,\text{kin}}(b_\perp)$, and it indicates that the difference, albeit small, is induced by the presence of quark-gluon interactions.

\section{Summary and Conclusion}

In this work, we have investigated the spin-orbit correlation in the pion by going beyond the leading Fock-state approximation. Pion, being a spin-0 particle, doesn't allow total intrinsic spin or orbital angular momentum of quarks. Nevertheless, the dynamical nature of QCD allows for non-vanishing correlation between the spin and orbital angular momentum of quarks. Previous studies of this correlation have largely been limited to the leading Fock sector. However, it is important to study the non-trivial effect of gluons on the spin-orbit correlation of quarks. In consideration with this, we constructed a pion model which incorporates gluonic degree of freedom through the inclusion of the the higher Fock component of the LFWF.  The three-particle LFWF was modeled using the quark-gluon-quark interaction in light-front QCD Hamiltonian. These three-particle LFWFs are expressed in terms of the two-particle LFWFs and perturbative vertices corresponding to gluon emission from the quark or antiquark. The LFWFs are Boost invariant  due to their dependence on relative momenta. 
The spin-independent part of the two-particle LFWFs is inspired by the AdS/QCD correspondence, and parameters are determined by fitting the $\bar{u}$ quark unpolarized PDF of the negative pion with the global QCD analysis results of the JAM 2021 collaboration at $Q=1$ GeV.
Using the model, we have predicted the results for quark chiral even GPD for both $ q \bar{q}$ and $ q \bar{q} g$ sectors while observing the chiral odd GPD contribution is negligible. Using the first and second moments of chiral even GPD, we have analyzed the results for electromagnetic form factor and momentum gravitational form factor along with their sum rules. 

Using the overlap representations of GPD ($F_1(x,0,t)$ and $H_1(x,0,t)$) and GTMD ($G_{1,1}$), we have calculated the kinetic and canonical quark longitudinal SOC, respectively. We show that with the inclusion of a gluon in the pion state, the GPD and the GTMD approach do not give  same result for the SOC.  
We also compute the pdf of longitudinal SOC through both approaches and find that it is completely negative when computed via the GTMD; however, it becomes positive at low values of $x$ when computed using GPDs overlaps. We observe the dependence of $C^q_{\text{GPD}}$ and $C^q_{\text{GTMD}}$ as a function of momentum transfer squared and find that it is negative in the whole region in both approaches. The integrated SOC number from $C^q_{\text{GPD}}$ and $C^q_{\text{GTMD}}$ are both negative, suggesting that quark spin and orbital angular momentum tend to be anti-correlated. We present the results for the $x$-dependence and the $k_\perp$-dependence of the longitudinal spin-orbit correlation, which differ, as expected,  as GPD and GTMD have different dependence on $k_\perp$. We have also obtained the spatial distribution of SOC and observe that the distribution is Gaussian like which peaks around $ b_\perp \approx 0.25 $ and then decreases monotonically. 
Along with a comparative study of the kinetic and canonical spin–orbit correlations (SOCs), we have highlighted the difference between the two. {{ In the literature, the difference between the kinetic and canonical SOC has been investigated  and interpreted in terms of higher-twist quark–gluon correlators and  gauge links.  As shown in Ref.~\cite{Hatta:2024otc}, the canonical SOC can be decomposed into higher-twist distributions that encode explicit quark–gluon interactions in the operator structure. While  higher-twist contributions are essential for a complete description of the correlation, here we show that inclusion of the higher Fock components of the pion LFWF also brings in quark-gluon dynamics and influence the SOC. In addition,}}  we focus on the longitudinal spin–orbit correlation (SOC) and orbital angular momentum of quarks where parton helicity, projection of OAM and their correlation is along the direction of motion. This study can be extended to the transverse direction where one can investigate {how transverse components of quark spin } and orbital {angular momentum interact with each other}. The definitions and decompositions in transverse directions are not as fairly known as they are explored for the longitudinal direction particularly in the context of transverse spin-orbit-correlation. Quark transverse spin–orbit correlations have been defined and analyzed in terms of the quark kinetic energy–momentum tensor and transversity decomposition of angular momentum. In the nucleon, it can be defined through the correlation between quark transversity and quark OAM as discussed in ~\cite{Bhoonah:2017olu}. 
Another next steps are the exploration of the corresponding gluonic quantities, especially in light of the scientific goals of the upcoming Electron–Ion Collider, where gluon dynamics will play a central role. The present model can be modified to study direct gluonic { contributions} to the pion. This study will complete the momentum sum rule in terms of quark and gluon SOC introduced in~\cite{Hatta:2024otc}.

\section{Acknowledgment}
PC thanks the organizers and participants of Joint International Workshop on Hadron Structure and Spectroscopy (IWHSS 2025) and the QCD Structure of the Nucleon (QCD-N'25) for helpful feedback during the presentation of this work.
\appendix
\section{Three Particle LFWFs}\label{3lfwf}
In this section, we present the analytic forms of three particle LFWFs as discussed in ~\cite{Harindranath:1998pd,Harindranath:1996sj}

\begin{eqnarray}
	\psi_3^{\uparrow \uparrow \lambda_1}=&A(x_1,x_2) \Bigg[ \chi_{\uparrow}^{\dagger} \big [ { 2 k_3^\perp \over 1 - x_1 -x_2} - { (\sigma^\perp. k_1^\perp
		- i m) \over x_1} \sigma^\perp + \sigma^\perp {(\sigma^\perp. k_2^\perp -
		im) \over 1-x_2} \big] \left( \chi_{\uparrow}. (\epsilon^\perp_{\lambda_1})^* \psi_2^{\uparrow, \uparrow} + \chi_{\downarrow}. (\epsilon^\perp_{\lambda_1})^* \psi_2^{\downarrow, \uparrow} \right) \nonumber \\
	&  - \chi_{\downarrow}^{\dagger}	 \big [ { 2 k_3^\perp \over 1 - x_1 -x_2} - \sigma^\perp
	{ (\sigma^\perp. k_2^\perp
		- i m) \over x_2}  +  {(\sigma^\perp. k_1^\perp -
		im) \over 1-x_1} \sigma^\perp 
	\big]  \left( \chi_{\uparrow}. (\epsilon^\perp_{\lambda_1})^* \psi_2^{\uparrow, \downarrow} + \chi_{\downarrow}. (\epsilon^\perp_{\lambda_1})^* \psi_2^{\uparrow, \uparrow} \right) \Bigg]
\end{eqnarray}
\begin{eqnarray}
	\psi_3^{\uparrow \uparrow \lambda_1 *}=&A(x,x_2) \Bigg[\left( \chi_{\uparrow }^{\dagger}. (\epsilon^\perp_{\lambda_1}) \psi_2^{\uparrow, \uparrow *} + \chi_{\downarrow}^{\dagger} (\epsilon^\perp_{\lambda_1}) \psi_2^{\downarrow, \uparrow *} \right)  \big [ { 2 k_3^\perp \over 1 - x_1 -x_2} -\sigma^\perp  { (\sigma^\perp. k_1^\perp
		+ i m) \over x_1} +  {(\sigma^\perp. k_2^\perp +
		im) \over 1-x_2} \sigma^\perp \big] \chi_{\uparrow} \nonumber \\
	&  - \chi_{\downarrow}^{\dagger}	 \big [ { 2 k_3^\perp \over 1 - x_1 -x_2} - \sigma^\perp
	{ (\sigma^\perp. k_2^\perp
		- i m) \over x_2}  +  {(\sigma^\perp. k_1^\perp -
		im) \over 1-x_1} \sigma^\perp 
	\big]  \left( \chi_{\uparrow}. (\epsilon^\perp_{\lambda_1})^* \psi_2^{\uparrow, \downarrow} + \chi_{\downarrow}. (\epsilon^\perp_{\lambda_1})^* \psi_2^{\uparrow, \uparrow} \right) \Bigg]
\end{eqnarray}

\begin{eqnarray}
	\psi_3^{\downarrow \downarrow \lambda_1}=&A(x_1,x_2) \Bigg[ \chi_{\downarrow}^{\dagger} \big [ { 2 k_3^\perp \over 1 - x_1 -x_2} - { (\sigma^\perp. k_1^\perp
		- i m) \over x_1} \sigma^\perp + \sigma^\perp {(\sigma^\perp. k_2^\perp -
		im) \over 1-x_2} \big] \left( \chi_{\uparrow}. (\epsilon^\perp_{\lambda_1})^* \psi_2^{\uparrow, \downarrow} + \chi_{\downarrow}. (\epsilon^\perp_{\lambda_1})^* \psi_2^{\downarrow, \downarrow} \right) \nonumber \\
	&  - \chi_{\uparrow}^{\dagger}	 \big [ { 2 k_3^\perp \over 1 - x_1 -x_2} - \sigma^\perp
	{ (\sigma^\perp. k_2^\perp
		- i m) \over x_2}  +  {(\sigma^\perp. k_1^\perp -
		im) \over 1-x_1} \sigma^\perp 
	\big]  \left( \chi_{\uparrow}. (\epsilon^\perp_{\lambda_1})^* \psi_2^{\downarrow, \downarrow} + \chi_{\downarrow}. (\epsilon^\perp_{\lambda_1})^* \psi_2^{\downarrow, \uparrow} \right) \Bigg]
\end{eqnarray}

\begin{eqnarray}
	\psi_3^{\uparrow \downarrow \lambda_1}=&A(x_1,x_2) \Bigg[ \chi_{\uparrow}^{\dagger} \big [ { 2 k_3^\perp \over 1 - x_1 -x_2} - { (\sigma^\perp. k_1^\perp
		- i m) \over x_1} \sigma^\perp + \sigma^\perp {(\sigma^\perp. k_2^\perp -
		im) \over 1-x_2} \big] \left( \chi_{\downarrow}. (\epsilon^\perp_{\lambda_1})^* \psi_2^{\downarrow, \downarrow} + \chi_{\uparrow}. (\epsilon^\perp_{\lambda_1})^* \psi_2^{\uparrow, \downarrow} \right) \nonumber \\
	&  - \chi_{\uparrow}^{\dagger}	 \big [ { 2 k_3^\perp \over 1 - x_1 -x_2} - \sigma^\perp
	{ (\sigma^\perp. k_2^\perp
		- i m) \over x_2}  +  {(\sigma^\perp. k_1^\perp -
		im) \over 1-x_1} \sigma^\perp 
	\big]  \left( \chi_{\uparrow}. (\epsilon^\perp_{\lambda_1})^* \psi_2^{\uparrow, \downarrow} + \chi_{\downarrow}. (\epsilon^\perp_{\lambda_1})^* \psi_2^{\uparrow, \uparrow} \right) \Bigg]
\end{eqnarray}

\begin{eqnarray}
	\psi_3^{\downarrow \uparrow \lambda_1}=&A(x_1,x_2) \Bigg[ \chi_{\downarrow}^{\dagger} \big [ { 2 k_3^\perp \over 1 - x_1 -x_2} - { (\sigma^\perp. k_1^\perp
		- i m) \over x_1} \sigma^\perp + \sigma^\perp {(\sigma^\perp. k_2^\perp -
		im) \over 1-x_2} \big] \left( \chi_{\downarrow}. (\epsilon^\perp_{\lambda_1})^* \psi_2^{\downarrow, \uparrow} + \chi_{\uparrow}. (\epsilon^\perp_{\lambda_1})^* \psi_2^{\uparrow, \uparrow} \right) \nonumber \\
	&  - \chi_{\downarrow}^{\dagger}	 \big [ { 2 k_3^\perp \over 1 - x_1 -x_2} - \sigma^\perp
	{ (\sigma^\perp. k_2^\perp
		- i m) \over x_2}  +  {(\sigma^\perp. k_1^\perp -
		im) \over 1-x_1} \sigma^\perp 
	\big]  \left( \chi_{\uparrow}. (\epsilon^\perp_{\lambda_1})^* \psi_2^{\downarrow, \downarrow} + \chi_{\downarrow}. (\epsilon^\perp_{\lambda_1})^* \psi_2^{\downarrow, \uparrow} \right) \Bigg]
\end{eqnarray}
where \begin{eqnarray}
	A(x_1,x_2)=	{ 1 \over E} (-) { g \over \sqrt{2 (2 \pi)^3}} T^a
	{ 1 \over \sqrt{1 - x_1 - x_2}}
\end{eqnarray}
with 
\begin{eqnarray}
	E=
	\big[ M^2  - {m^2 + (k_1^\perp)^2 \over x_1} -
	{m^2 + (k_2^\perp)^2 \over x_2} - {(k_3^\perp)^2 \over 1 - x_1 -x_2}
	\big ],
\end{eqnarray}

\section{LFWF Overlap representation }
In this section, we present the expressions of LFWF overlaps of the quantities of interest for this work. These overlaps have been calculated by sandwiching the operator definitions of the GPDS, GTMDs, and spin-orbit correlation given in Sec. III. All the distributions are evaluated at zero skewness throughout this work. 
\subsection{GPD overlaps}
Using Eq. \eqref{H}, we find the LFWF overlap for the chiral even GPD as
\begin{align}
    F_1(x_1,0,t) &= \frac{1}{16\pi^3}\sum_{\sigma_1,\sigma_2,\lambda_3} \int \int \int dx_2 d^2q^\perp_1 d^2q^\perp_2 \left[\psi_2^{*\sigma_1,\sigma_2}(x_1,q'^\perp_1)\psi_2^{\sigma_1,\sigma_2}(x_1,q^\perp_1) \,\,+\right. \nonumber \\
    & \left. \hspace{6cm}\psi_3^{*\sigma_1,\sigma_2,\lambda_3}(x_1,x_2,q'^\perp_1,q'^\perp_2)\psi_3^{\sigma_1,\sigma_2,\lambda_3}(x_1,x_2,q^\perp_1,q^\perp_2) \right]\label{Hlfwf}
\end{align}
where $x_1=\dfrac{k^+}{P^+}, \,\,q'^\perp_1 = q^\perp_1 + (1-x_1) \Delta^\perp, \,\,q'^\perp_2= q^\perp_2-x_2\Delta^\perp$. 
Similarly, the three particle LFWF representation of chiral odd GPD for zero skewness can be obtained by using Eq. \eqref{ET}
\begin{align}
    \frac{i \Delta^{j}}{2M_\mathcal{P}}H_1(x_1,0,t) &= \frac{-1}{16\pi^3}\sum_{\sigma_2,\lambda_3} \int \int \int dx_2 d^2q^\perp_1 d^2q^\perp_2 \left\{(-i)^j\left[\psi_2^{*\uparrow,\sigma_2}\psi_2^{\downarrow,\sigma_2}+\psi_3^{*\uparrow,\sigma_2,\lambda_3}   \psi_3^{\downarrow,\sigma_2,\lambda_3}\right] \,\, + \right. \nonumber \\
    &\left. \hspace{7cm} (i)^j \left[\psi_2^{*\downarrow,\sigma_2}\psi_2^{\uparrow,\sigma_2}+\psi_3^{*\downarrow,\sigma_2,\lambda_3}   \psi_3^{\uparrow,\sigma_2,\lambda_3}\right]\right\}
\end{align}
where $j=1,2$. We omit the functional dependence of LFWFs as they are the same as in Eq.\eqref{Hlfwf}. The two components can be further combined to give a more compact expression for $E_T$ as follows
\begin{align}
    \frac{\Delta^*_\perp}{2M_\mathcal{P}}H_1(x_1,0,t) &= -\frac{1}{8\pi^3}\sum_{\sigma_2,\lambda_3} \int \int \int dx_2 d^2q^\perp_1 d^2q^\perp_2 \left[\psi_2^{*\downarrow,\sigma_2}\psi_2^{\uparrow,\sigma_2}+\psi_3^{*\downarrow,\sigma_2,\lambda_3}   \psi_3^{\uparrow,\sigma_2,\lambda_3}\right]
\end{align}
or 
\begin{align}
    \frac{\Delta_\perp}{2M_\mathcal{P}}H_1(x_1,0,t) &= \frac{1}{8\pi^3}\sum_{\sigma_2,\lambda_3} \int \int \int dx_2 d^2q^\perp_1 d^2q^\perp_2 \left[\psi_2^{*\uparrow,\sigma_2}\psi_2^{\downarrow,\sigma_2}+\psi_3^{*\uparrow,\sigma_2,\lambda_3}   \psi_3^{\downarrow,\sigma_2,\lambda_3}\right] \label{Elfwf}
\end{align}
where $ \Delta_\perp $ is momentum transfer. 
\subsection{GTMD overlaps}
The 3-particle LFWF representation of the quark GTMD is given as 
\begin{align}
     -\frac{i \epsilon^{i j} k^i \Delta^{j}}{M^2}G_{1,1}^\mathcal{\pi}(x_1,0,k_\perp,\Delta_\perp) &= \frac{1}{16\pi^3} \sum_{\sigma_2,\lambda_3}\int dx_2 d^2q_2^\perp \left\{ \left[\psi_2^{*\uparrow,\sigma_2}\psi_2^{\uparrow,\sigma_2}+\psi_3^{*\uparrow,\sigma_2,\lambda_3}\,\psi_3^{\uparrow,\sigma_2,\lambda_3}\right]- \right. \nonumber \\
     & \left. \hspace{6cm} \left[\psi_2^{*\downarrow,\sigma_2}\psi_2^{\downarrow,\sigma_2} +\psi_3^{*\downarrow,\sigma_2,\lambda_3}\,\psi_3^{\downarrow,\sigma_2,\lambda_3}\right]\right\}
\end{align}

\subsection{Analytic expression of SOC from two particle only}
\label{twoParGPDGTMD}
The SOC in kinetic decomposition can be written in terms of GPDs. Here, we will focus only the two particle overlap and their contribution to corresponding SOC. 
\begin{align}
C^{q/\pi}_z\big{|}_{\textrm{GPD}} &= \frac{1}{2} \int_{-1}^1 dx\,  \left[- F_1^{\mathcal{\pi}}(x,0,t) +\frac{m_q}{M} H_1^{\mathcal{\pi}}(x,0,t) \right] \nonumber\\
      &= \int dx d^2\bm k_\perp{4(1-x)m_q^2-2(\bm k_\perp^2-(1-x)^2 \frac{\Delta_\perp^2}{4}+m_q^2)\over 32\pi^3(\sqrt{(\mathbf{k}_\perp +(1-x)\frac{\Delta_\perp}{2})^2+m_q^2)} \sqrt{(\mathbf{k}_\perp -(1-x)\frac{\Delta_\perp}{2})^2+m_q^2)}} \nonumber \\ 
      & \varphi^*_\pi(x,\mathbf{k}_\perp +(1-x)\frac{\Delta_\perp}{2}) \,  \varphi_\pi(x,\mathbf{k}_\perp -(1-x)\frac{\Delta_\perp}{2}) \label{eq:modelGPD}
\end{align}
We can further simplify the above integral in the forward limit $ \Delta=0$, 
\begin{align}
C^{q/\pi}_z\big{|}_{\textrm{GPD}} &= \int dx d^2\bm k_\perp{4(1-x)m_q^2-2(\bm k_\perp^2+m_q^2)\over 32\pi^3  (\mathbf{k}_\perp^2+m_q^2)} |\varphi_\pi(x,\mathbf{k}_\perp)|^2 \, \nonumber \\
& =\int dx \left[\frac{2 m^{2}}{x \kappa^{2}} \,
\Gamma\!\left(0,\, 2 m^{2} (1 - x)^{-\beta} x^{-\alpha} /\kappa^{2}\right)
- \frac{
e^{-2 m^{2} (1 - x)^{-\beta} x^{-\alpha}/\kappa^{2}}
}{
2  (1 - x)^{1 - \beta} x^{1-\alpha}
} \right]
\end{align}
Likewise, we can also evaluate the SOC from canonical decomposition which is related to $ G_{1,1}$ GTMD:
 \begin{align}
	C^q_z|_{GTMD} &= \int dx\, d^2k_\perp \, \frac{k_\perp^2}{M_\pi^2} \,
	G_{1,1}(x,k_\perp,\Delta_\perp,k_\perp.\Delta_\perp) \nonumber \\
    & = -\int dx d^2\bm k_\perp{(1-x)\bm k_\perp^2\over 8\pi^3} \frac{\varphi^*_\pi(x,\mathbf{k}_\perp +(1-x)\frac{\Delta_\perp}{2}) \,  \varphi_\pi(x,\mathbf{k}_\perp -(1-x)\frac{\Delta_\perp}{2})}{\sqrt{(\mathbf{k}_\perp +(1-x)\frac{\Delta_\perp}{2})^2+m_q^2)}\sqrt{(\mathbf{k}_\perp -(1-x)\frac{\Delta_\perp}{2})^2+m_q^2)}}
\end{align} 
In the forward limit:
\begin{align}
	C^q_z|_{GTMD} &= -\int dx d^2\bm k_\perp{(1-x)\bm k_\perp^2\over 8\pi^3} \frac{|\varphi_\pi(x,\mathbf{k}_\perp)|^2}{(\mathbf{k}_\perp^2+m_q^2)} \nonumber \\
    &= \int dx \left[\frac{2 m^{2}}{x \kappa^{2}}\,
\Gamma\!\Big(0,\,2 m^{2}(1-x)^{-\beta} x^{-\alpha} /\kappa^{2}\Big) -\frac{e^{-2 m^{2}(1-x)^{-\beta} x^{-\alpha} /\kappa^{2}}}{  (1-x)^{-\beta} x^{1-\alpha} } \right]
\end{align} 
The difference in above two distributions $C^q_{GPD}(x) $ and $C^q_{GTMD}(x)$: 
\begin{align}
C^q_z|_{GPD}-C^q_z|_{GTMD} &= 
\int dx \left[ \frac{e^{-2 m^{2}(1-x)^{-\beta} x^{-\alpha} /\kappa^{2}}}{2  (1-x)^{1-\beta} x^{1-\alpha} }\left(1-2x\right ) \right]
\end{align} 
The difference is coming because of the choice of $ \phi(x,k_\perp)$. In Ref.~\cite{Tan:2021osk} in equation (41) of that work, they used a relation in the integral which helps in the matching of GPD and GTMD results, but that integral relation is valid for their choice of $ \phi(x,k_\perp) $
\bibliography{ref.bib}

\end{document}